%% file: main.tex
\documentclass[sigconf, nonacm]{acmart}
\AtBeginDocument{%
  }

\usepackage{hyperref}
\usepackage{booktabs}
\newcommand{\ra}[1]{\renewcommand{\arraystretch}{#1}}
\usepackage{subcaption}

\input{include/packages}

\input{include/macros}
\input{include/colors}

\newcommand{\mycomment}[3]{}

\newcommand{\boldparagraph}[1]{\paragraph{\upshape\bfseries #1}} %

\usepackage{xspace}
\newcommand{\tool}{\textsf{DARSY}\xspace}

\begin{document}

\title[\textsf{DARSY}: A Tool for Data-Driven Abstraction and Policy Synthesis for Stochastic Systems with Unknown Dynamics]{\tool: A Tool for Data-Driven Abstraction and Policy Synthesis for Stochastic Systems with Unknown Dynamics \texorpdfstring{\vspace{5mm}}{}%
}

\author{Thom Badings}
\authornote{Equal contribution}
\affiliation{%
  \institution{RWTH Aachen University}
  \city{Aachen}
  \country{Germany}
}
\email{thom.badings@cs.rwth-aachen.de}

\author{Ibon Gracia}
\authornotemark[1]
\affiliation{%
  \institution{University of Colorado Boulder}
  \city{Boulder, Colorado}
  \country{USA}
}
\email{ibon.gracia@colorado.edu}

\author{Mahdi Nazeri}
\authornotemark[1]
\affiliation{%
 \institution{University of Oxford \& MPI SWS}
 \city{Oxford}
 \country{United Kingdom}
 }
\email{mahdi.nazeri@cs.ox.ac.uk}

\author{Morteza Lahijanian}
\affiliation{%
  \institution{University of Colorado Boulder}
  \city{Boulder, Colorado}
  \country{USA}
}
\email{morteza.lahijanian@colorado.edu}

\author{Alessandro Abate}
\affiliation{%
  \institution{University of Oxford}
  \city{Oxford}
  \country{United Kingdom}
}
\email{alessandro.abate@cs.ox.ac.uk}

\begin{abstract}
This paper presents \tool: a software tool for abstraction-based verification and policy synthesis in discrete-time stochastic control systems with unknown dynamics and noise distributions.
\tool requires only a simulator of the system and a Lipschitz bound on its dynamics, and needs no knowledge of dynamics or noise distributions.
The tool is fully automated and constructs a finite interval Markov decision process (IMDP) abstraction from sampled transitions, whose probability intervals are obtained via statistical techniques and Lipschitz reasoning.
Robust dynamic programming on this IMDP then yields a policy together with a lower bound on the probability that the unknown system satisfies a desired specification.
This lower bound holds with a user-specified confidence; that is, it is probably approximately correct (PAC).
The tool is implemented in Python and supports general Lipschitz-continuous system dynamics and specifications expressed in syntactically co-safe LTL (scLTL), which includes reach-avoid and finite-horizon safety specifications as special cases. 
Beyond standard IMDPs, \tool implements a two-layer IMDP (2IMDP) abstraction that also bounds the probability of transitioning to aggregated state blocks, yielding smaller abstraction errors than IMDP abstractions and, as a result, tighter satisfaction guarantees.
We demonstrate \tool's broad applicability across a variety of benchmarks and specifications.
\end{abstract}

\ccsdesc[500]{Theory of computation~Markov decision processes}
\ccsdesc[500]{Theory of computation~Abstraction}
\ccsdesc[500]{Mathematics of computing~Markov processes}
\ccsdesc[500]{Computing methodologies~Planning under uncertainty}

\keywords{Stochastic dynamical systems, formal policy synthesis, data-driven abstraction, interval Markov decision processes, PAC guarantees}

\maketitle

\section{Introduction}
\label{sec:introduction}
The dynamics of many control systems, such as unmanned drones, mobile robots, and thermodynamic processes, are modeled by \emph{nonlinear stochastic difference equations}~\cite{soderstrom2002discrete,mesbah2016stochastic}. 
These systems often must satisfy strict requirements despite their stochastic dynamics.
For example, a mobile robot may need to visit a sequence of target locations while avoiding obstacles, despite random disturbances such as wind or wheel slip.
Such requirements, called \emph{specifications}, are expressed formally in temporal logics, such as \emph{syntactically co-safe linear temporal logic} (scLTL)~\cite{DBLP:journals/fmsd/KupfermanV01,belta2017formal}, which includes the ubiquitous \emph{reach-while-avoid} task as a special case~\cite{DBLP:conf/cav/FanMM018,DBLP:journals/automatica/SummersL10}.
Because of the randomness in the system's dynamics, such a specification generally cannot be satisfied with certainty.
The goal of \emph{formal policy synthesis}~\cite{belta2017formal} is therefore to compute a control policy that achieves a high probability of satisfying the specification, together with guaranteed (and tight) bounds on this probability.

A well-established approach to formal policy synthesis is to abstract the system into a finite Markov decision process (MDP) or variant thereof~\cite{LSAZ21,DBLP:journals/tac/LahijanianAB15,DBLP:journals/jair/BadingsRAPPSJ23,DBLP:journals/tac/HaesaertS21,DBLP:journals/corr/abs-2507-02213}.
One can then synthesize an optimal \emph{strategy} on this abstraction and refine it into a policy for the original system, such that guarantees on the abstraction carry over to the system~\cite{DBLP:journals/automatica/AbatePLS08,DBLP:books/daglib/0032856,DBLP:conf/hybrid/MathiesenHL25}.
A broad range of tools implement such approaches, including FAUST$^2$~\cite{FAUST15}, StocHy~\cite{StocHy19}, AMYTISS~\cite{AMYTISS20}, SySCoRe~\cite{Huijgevoort2023SySCoRe}, and IMPaCT~\cite{IMPaCT24}.
Several of these tools are compared in the ARCH-Competition on stochastic models~\cite{DBLP:journals/corr/abs-2506-17602}.
However, all of these tools are \emph{model-based}: they require an explicit and exact model of the dynamics, often including the distribution of the stochastic noise.
In practice, such a model is often unavailable because the dynamics may be too complex to model accurately, or the system may be accessible only through a simulator or experimental data.

\emph{Data-driven abstractions} address this limitation by constructing abstractions from sampled transitions of the system, with guarantees that typically hold with a user-specified confidence~\cite{DBLP:conf/adhs/MakdesiGF21,DBLP:journals/csysl/CoppolaPM23,lavaei2023compositional,DBLP:journals/corr/abs-2206-08069,DBLP:journals/automatica/HashimotoSKUD22,DBLP:conf/cdc/DevonportSA21,DBLP:conf/cdc/BanseRAJ23,DBLP:journals/csysl/PeruffoM23,schon2024data,DBLP:conf/l4dc/GraciaBLL24,DBLP:conf/hybrid/JacksonLFL21,DBLP:journals/jair/BadingsRAPPSJ23,DBLP:journals/csysl/LavaeiSFZ23, DBLP:journals/csysl/AjeleyeLZ23}.
Some approaches only assume the stochastic noise distribution to be unknown~\cite{DBLP:journals/corr/abs-2412-11343,DBLP:conf/l4dc/GraciaBLL24,DBLP:conf/aaai/BadingsRA023,DBLP:conf/aaai/BadingsA00PS22}, while others assume completely unknown dynamics with only a known Lipschitz constant~\cite{DBLP:conf/l4dc/NazeriBSA25,DBLP:conf/cdc/NazeriBSSA25}.
Yet, despite this growing body of literature, tool support for data-driven abstractions is lacking.
To the best of our knowledge, no general-purpose software tool exists for data-driven abstraction and formal policy synthesis for discrete-time stochastic systems.
As a result, applying the developed methods to a new system requires substantial implementation effort, which limits adoption of the results and further development of data-driven abstraction-based policy synthesis.

\subsection*{\tool: a tool for data-driven abstraction and policy synthesis}
We present \tool, a software tool for data-driven abstraction and policy synthesis for stochastic systems with unknown~dynamics.
The tool is implemented in Python with JAX~\cite{jax2018github}, enabling just-in-time compilation and vectorized computation of the abstraction and policy.
\tool requires only a black-box, resettable simulator of the system and a Lipschitz constant of its dynamics, and needs no knowledge of the dynamics or the noise distribution.

Lipschitz continuity is a standard assumption in control theory~\cite{khalil2002nonlinear} and is central to how \tool extrapolates from observed transitions~\cite{DBLP:journals/tac/AlanwarKAJ23}.
Such knowledge about the regularity of the dynamics is needed for a sound abstraction from finitely many observations (see~\cref{sec:algorithms}).
In practice, Lipschitz constants can be obtained from expert knowledge or worst-case reasoning in parametric models.
Given a prior over the dynamics, statistical techniques~\cite{casella2002statistical} yield confidence intervals for the Lipschitz constant.
If only observations are available, data-driven methods can robustly estimate Lipschitz constants under additional smoothness assumptions~\cite{DBLP:conf/aistats/ZhangMSS24,DBLP:journals/tmlr/HuangRC23}.

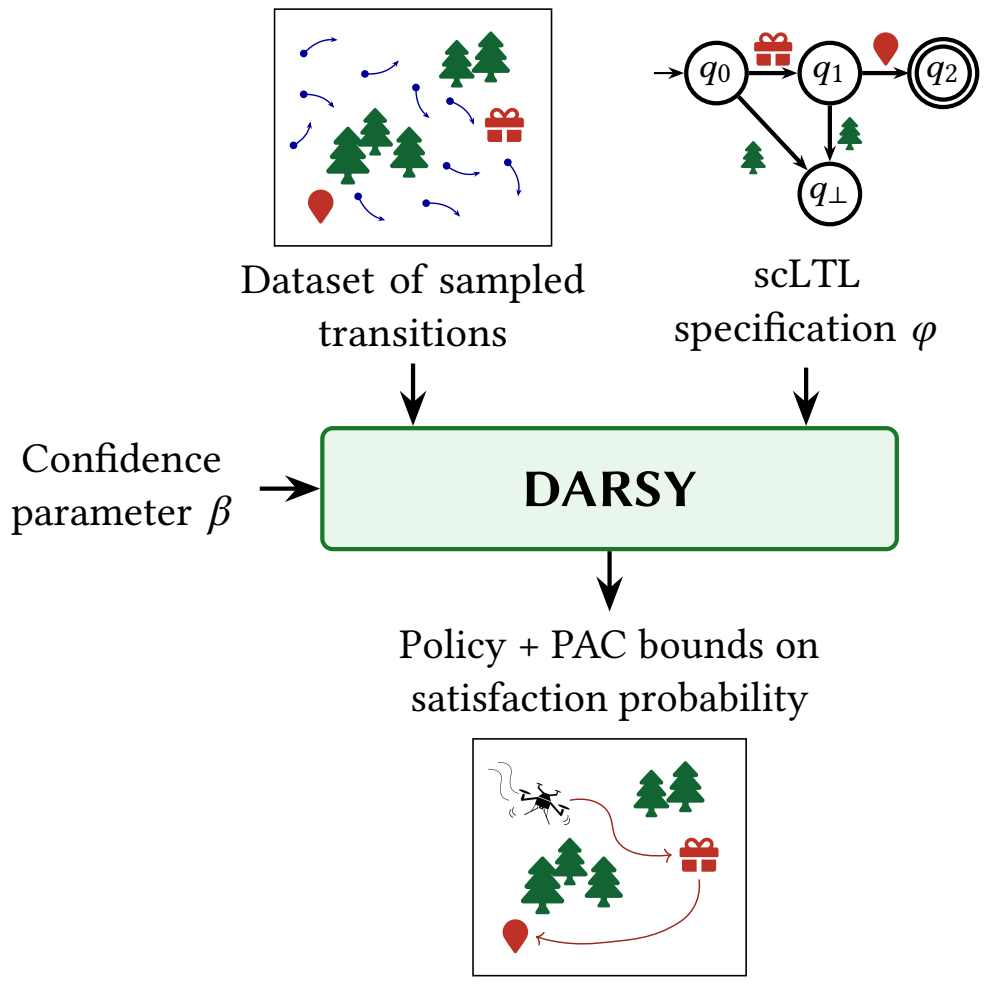
\begin{figure}[t]
    \centering
    \def\boxradius{2pt}
    \newsavebox{\introsamplesbox}
    \sbox{\introsamplesbox}{\scalebox{0.45}{\input{figures/intro_samples}}}
    \newsavebox{\introautomatonbox}
    \sbox{\introautomatonbox}{\input{figures/intro_automaton}}
    \newsavebox{\introtrajectorybox}
    \sbox{\introtrajectorybox}{\scalebox{0.45}{\input{figures/intro_trajectory}}}
    \begin{tikzpicture}[
        font=\small,
        every node/.style={align=center},
        io/.style={text width=2.3cm},
        toolbox/.style={draw, thick, rounded corners=\boxradius, minimum height=8mm},
        arrow/.style={-{Stealth}, thick},
    ]
        \node[io] (data) at (-1.3,0) {Dataset of sampled \\ transitions};
        \node[io] (spec) at (1.3,0) {scLTL\\specification $\varphi$};
        \node[draw=black, line width=0.3pt, inner sep=1mm, above=0mm of data] (samples) {\usebox{\introsamplesbox}};
        \node[inner sep=0pt, above=1mm of spec] (automaton) {\usebox{\introautomatonbox}};

        \node[toolbox, below=4mm of {$(data.south)!0.5!(spec.south)$}, minimum width=3.8cm, font=\large, draw=darkgreen!75!black, fill=darkgreen!10] (tool) {\textbf\tool};
        \foreach \n in {data, spec}
            \draw[arrow] (\n.south) -- (\n |- tool.north);

        \node[io, text width=1.6cm, left=4mm of tool] (beta) {Confidence\\parameter $\beta$};
        \draw[arrow] (beta) -- (tool);
    
        \node[below=4mm of tool] (out) {Policy + PAC bounds on \\ satisfaction probability};
        \draw[arrow] (tool) -- (out);
        \node[draw=black, line width=0.3pt, inner sep=1mm, below=0mm of out] (trajectory) {\usebox{\introtrajectorybox}};
    \end{tikzpicture}
    
    \caption{Based on sampled transitions (shown in blue) from a discrete-time stochastic system, \tool computes a policy together with PAC bounds on the probability that the system satisfies a given scLTL specification (shown as automaton).}
    \label{fig:overview_simple}
\end{figure}

\paragraph{Tool overview}
As shown in \cref{fig:overview_simple}, \tool uses sampled transitions from this simulator to construct an \emph{interval MDP} (IMDP) abstraction~\cite{DBLP:journals/ai/GivanLD00,DBLP:journals/ior/NilimG05,DBLP:journals/mor/Iyengar05}, whose transition probability intervals are derived from the samples via Clopper-Pearson confidence intervals and the Lipschitz constant~\cite{DBLP:conf/cdc/NazeriBSSA25}.
These imprecise transition probabilities account for discretization errors in the abstraction, which is thus sound and captures all behaviors of interest in the unknown system.
In parallel, \tool translates the scLTL specification into a \emph{deterministic finite automaton} (DFA) using Spot~\cite{DBLP:conf/cav/Duret-LutzRCRAS22}.
Robust dynamic programming on the product of the IMDP and the DFA then yields an abstract strategy, which is refined into a finite-memory policy for the unknown system.
This policy comes with formal bounds on the probability that the system satisfies the specification, which hold with a user-specified confidence; that is, the bound is \emph{probably approximately correct} (PAC).
Beyond standard IMDPs, \tool also implements a data-driven variant of the two-layer IMDP (2IMDP) abstraction from~\cite{DBLP:journals/corr/abs-2507-02213}, which additionally bounds the probability of transitioning into aggregated blocks of states and thereby yields tighter guarantees on synthesized policies.

\paragraph{Benchmarks}
We evaluate \tool on an array of linear and nonlinear stochastic system benchmarks from the literature~\cite{DBLP:conf/cdc/NazeriBSSA25,DBLP:conf/nips/TowersKBCDGKKKP25,DBLP:journals/corr/abs-2507-02213}.
We synthesize policies for different scLTL, reach-avoid, and safety specifications.
The evaluation shows the tool's ability to synthesize policies with tight bounds on the satisfaction probabilities.

\paragraph{Contributions}
Our main contributions are as follows:
\begin{enumerate}
    \item We present \tool, a Python tool that synthesizes policies with PAC guarantees for stochastic systems with unknown dynamics against scLTL specifications.
    The tool supports both IMDP and 2IMDP abstractions, and is implemented in JAX, enabling just-in-time compilation and vectorized computation of abstractions and policies.
    \item \tool integrates the data-driven abstraction and synthesis algorithms from~\cite{DBLP:conf/cdc/NazeriBSSA25} with a data-driven extension of~\cite{DBLP:journals/corr/abs-2507-02213}, yielding tighter abstractions and guarantees on~policies.
    \item We demonstrate the applicability of \tool on a set of benchmarks with different dynamics and specifications.
    We emphasize the generality of the tool to unknown, nonlinear dynamics and temporal logic specifications in scLTL.
\end{enumerate}

\paragraph{Outline}
The remainder of the paper is organized as follows.
In \cref{sec:theory}, we introduce the technical background and define the problem we solve.
\Cref{sec:algorithms} describes the structure of \tool and its algorithms, and \cref{sec:usage} explains how to use the tool.
Finally, in \cref{sec:experiments}, we evaluate \tool on a set of benchmarks.

\paragraph{Getting started with \tool}
The tool and its Python source code, as well as all benchmarks presented in~\cref{sec:experiments}, are available at
\url{https://github.com/tbadings/darsy}.

\section{Theoretical Underpinnings}
\label{sec:theory}

In this section, we define the discrete-time stochastic systems, specifications, and data-driven policy synthesis problem we study.

\paragraph{Notation}
We write the power set (the set of all subsets) over $X$ as~$2^X$.
For the natural numbers, we write $\Nat_{\geq 0} \coloneqq \{0\} \cup \Nat$.
A probability space $\tuple{\Omega, \mathcal{F}, \Prob}$ consists of a sample space $\Omega$, a $\sigma$-algebra $\mathcal{F}$, and a probability measure $\Prob \colon \mathcal{F} \to [0,1]$.
The set of all distributions over a set $X$ is written as $\distr{X}$.
For a set $Z$, we write $Z^*$ and $Z^\omega$ to denote the sets of finite and infinite sequences over $Z$, respectively.

\subsection{Discrete-time stochastic systems}
\label{subsec:dtss}

Consider a discrete-time stochastic dynamical control system $\system$, defined for all time steps $k \in \NN_{\geq 0}$ by the difference equation
\begin{equation}
    x_{k+1} = f(x_k, u_k, \noise_k), 
    \label{eq:system}
\end{equation}
where $x_k \in \X$ %
is the state at time $k$, defined on a continuous state space $\X \subseteq \Real^n$, and $u_k \in \U$ is the control input at time $k$, selected from a finite set of inputs $\U = \{u_1, \ldots, u_m\} \subset \RR^d$.
Moreover, $\{\noise_k\}_{k \in \NN_{\geq 0}}$ is a sequence of i.i.d. random variables $\noise_k \colon \Omega \to \Noise \subseteq \RR^p$ defined on the probability space $(\Omega, \mathcal{F}, \mathbb{P})$.
The dynamics function $f \colon \X \times \U \times \Noise \to \X$ is a general, nonlinear function that we assume to be measurable and Lipschitz continuous.
\begin{assumption}[Lipschitz continuity]
    \label{def:Lipschitz}
    The dynamics function $f \colon \X \times \U \times \Noise \to \X$ is $L_X$-Lipschitz continuous in $x$ (uniformly in $u$ and $\noise$), i.e., there exists a constant $L_X \in \mathbb{R}_{>0}$ such that
    \begin{equation*}
        |f(x_1, u, \noise) - f(x_2, u, \noise)| \leq L_X \cdot |x_1 - x_2|
    \end{equation*}
    for all $x_1, x_2 \in \X$, $u \in \U$, and $\noise \in \Noise$, with $|\cdot|$ being an appropriate norm on $\X$.
\end{assumption}

\paragraph{Policy}
Let $\PathsFin \coloneqq \X^*$ and $\PathsInf \coloneqq \X^\omega$ be the sets of finite and infinite \emph{paths} of the system $\system$, respectively.
The control inputs are selected by a policy $\policy$, defined as a measurable function $\policy \colon \PathsFin \to \U$, and the set of all policies is $\Pi^\system$.
Given a fixed policy $\pi\in\Pi^\system$ and initial state $x_0 \in \X$, the state of the closed-loop system evolves according to $x_{k+1} = f(x_k, \pi(x_0, \ldots, x_k), \noise_k)$, which induces a stochastic process over (infinite) paths $(x_0, x_1, \ldots)$.
We denote the induced probability measure associated with this stochastic process over paths by $\mathbb{P}_{\pi,x_0}^\system$~\cite{Bertsekas.Shreve78, DBLP:books/wi/Puterman94}, such that $\mathbb{P}_{\pi,x_0}^\system(\mathbf{X})$ is the probability that the closed-loop system generates a path in the measurable set of paths $\mathbf{X} \subseteq \PathsInf$.

\paragraph{Labeling function}
Let $\Y$ be a finite set of labels. 
The system $\system$ is equipped with a measurable labeling function $g \colon \X \to 2^{\Y}$ that maps each state to a (possibly empty) subset of $\Y$.
Thus, the label at time step $k$ is
\begin{equation}
    y_{k} = g(x_{k}).
    \label{eq:label}
\end{equation}
Although the theory (see~\cite{DBLP:conf/cdc/NazeriBSSA25,DBLP:journals/corr/abs-2412-11343}) that \tool implements is sound for every measurable labeling function, for ease of implementation and without much loss of generality, we assume that for all $\ell \in \Y$, the region $\{x\in \X : \ell \in g(x) \}$ can be partitioned into axis-aligned hyperrectangles. A sequence of labels $(y_0, y_1, \ldots)$ is called a \emph{trace}.
Since any path $(x_0, x_1, \ldots)$ corresponds with a unique trace $(g(x_0), g(x_1), \ldots)$, we can use the probability measure $\mathbb{P}_{\pi,x_0}^\system$ to assign the probability for any measurable set of traces $\mathbf{Y} \subseteq (2^\Y)^\omega$.

\subsection{Specifications}
\label{sec:specs}
We consider specifications expressed in \emph{syntactically co-safe linear temporal logic} (scLTL)~\cite{DBLP:journals/fmsd/KupfermanV01,belta2017formal}, the fragment of LTL~\cite{BaierKatoen08} in which negation is only applied to atomic propositions, and in which only the temporal operators \emph{next} ($\Next$) and \emph{until} ($\UUntil$) appear.
We take the set of \emph{atomic propositions} to be the labels $\Y$ of system $\system$. 

\begin{definition}[scLTL syntax]
    \label{def:scLTL}
    An \emph{scLTL formula} $\varphi$ over the set of labels $\Y$ is generated by the grammar
    \begin{equation*}
        \varphi \Coloneqq \True \mid \ell \mid \neg \ell \mid \varphi_1 \wedge \varphi_2 \mid \varphi_1 \vee \varphi_2 \mid \Next \varphi \mid \varphi_1 \UUntil \varphi_2,
    \end{equation*}
    where $\ell \in \Y$ is a label and $\varphi_1, \varphi_2$ are scLTL formulas.
\end{definition}

The semantics of scLTL are defined over infinite traces $\mathbf{y} = (y_0, y_1, \ldots) \in (2^\Y)^\omega$.
Let $\mathbf{y}^k \coloneqq (y_k, y_{k+1}, \ldots)$ be the suffix of $\mathbf{y}$ starting at time step $k \in \NN_{\geq 0}$.
The satisfaction relation $\models$ is defined inductively as $\mathbf{y}^k \models \True$ for all $k \in \NN_{\geq 0}$, and
\begin{align*}
    \mathbf{y}^k &\models \ell && \!\!\!\!\Leftrightarrow \ell \in y_k,
    \\
    \mathbf{y}^k &\models \neg \ell && \!\!\!\!\Leftrightarrow \ell \notin y_k,
    \\
    \mathbf{y}^k &\models \varphi_1 \wedge \varphi_2 && \!\!\!\!\Leftrightarrow \mathbf{y}^k \models \varphi_1 \text{ and } \mathbf{y}^k \models \varphi_2,
    \\
    \mathbf{y}^k &\models \varphi_1 \vee \varphi_2 && \!\!\!\!\Leftrightarrow \mathbf{y}^k \models \varphi_1 \text{ or } \mathbf{y}^k \models \varphi_2,
    \\
    \mathbf{y}^k &\models \Next \varphi && \!\!\!\!\Leftrightarrow \mathbf{y}^{k+1} \models \varphi,
    \\
    \mathbf{y}^k &\models \varphi_1 \UUntil \varphi_2 && \!\!\!\!\Leftrightarrow \exists j \geq k \colon \mathbf{y}^j \models \varphi_2 \text{ and } \mathbf{y}^i \models \varphi_1 \;\; \forall i \in \{k, \ldots, j-1\}.
\end{align*}
A trace $\mathbf{y}$ satisfies $\varphi$, written $\mathbf{y} \models \varphi$, if and only if $\mathbf{y}^0 \models \varphi$.
As usual, the \emph{eventually} operator is defined as $\Finally \varphi \coloneqq \True \UUntil \varphi$.
Common specifications include reachability ($\Finally \mathsf{goal}$), reach-avoid ($\neg \mathsf{unsafe} \UUntil \mathsf{goal}$), sequential reachability (e.g., $\Finally (\mathsf{goal}_1 \wedge \Finally \mathsf{goal}_2)$), and finite-horizon safety ($\bigwedge_{k=0}^{\horizon} \Next^k \mathsf{safe}$, where $\Next^k$ denotes $k$ nested next operators).

Every satisfying trace of an scLTL specification has a \emph{good prefix}, i.e., a finite prefix $(y_0, \ldots, y_k)$ such that every infinite extension of this prefix satisfies $\varphi$.
The set of good prefixes of an scLTL formula is a regular language, and there exists a \emph{deterministic finite automaton} (DFA)~\cite{DBLP:journals/fmsd/KupfermanV01,belta2017formal,DBLP:journals/ijrr/LacerdaFPH19} that accepts it.
As a result, the satisfaction of an scLTL formula is witnessed in finite time, enabling policy synthesis on the product of an abstraction with this DFA.

\begin{definition}[Satisfaction probability]
    \label{def:satprob_dyn}
    The probability that the system $\system$, with policy $\pi \in \Pi^\system$ and initialized at $x_0 \in \X$, satisfies the scLTL specification $\varphi$ is
    \begin{equation*}
        \satprob_{\pi,x_0}^\system(\varphi) \coloneqq \Prob_{\pi,x_0}^\system\big\{ \mathbf{y} \in (2^\Y)^\omega : \mathbf{y} \models \varphi \big\}.
    \end{equation*}
\end{definition}

Since the set of traces satisfying an scLTL formula is $\omega$-regular and thus measurable~\cite{BaierKatoen08}, and since $g$ is measurable, $\satprob_{\pi,x_0}^\system(\varphi)$ is well-defined.
Moreover, to maximize $\satprob_{\pi,x_0}^\system(\varphi)$, it suffices to consider \emph{finite-memory} policies~\cite{DBLP:journals/iandc/TkachevMKA17,BaierKatoen08}, which depend on the path $(x_0, \ldots, x_k)$ only through the current state $x_k$ and the state $q_k$ of the DFA after reading the trace prefix $(y_0, \ldots, y_k)$.

\subsection{Policy synthesis under unknown dynamics}
Given a system $\system$ and an scLTL specification $\varphi$, we aim to synthesize a policy $\pi$ with a high satisfaction probability $\satprob_{\pi,x_0}^\system(\varphi)$, together with certified bounds on this probability.
Crucially, we assume that the dynamics of the system $\system$ are unknown.
Instead, we assume only access to sampled state-input-state triples $(x, u, x^+)$, but the noise value $\noise(\xi)$, $\xi \in \Omega$, that produced each $x^+$ is unknown.
These samples are captured in the following dataset.

\begin{definition}[Dataset]
    \label{def:dataset}
    Let $x_i^+ \coloneqq f(x_i,u_i,\noise(\xi_i))$ be the successor of state $x_i$ under input $u_i$ and outcome $\xi_i \in \Omega$.
    A sampled dataset for system $\system$ of $N \in \NN$ state-input-state observations is defined as
    \begin{equation*}
    \Dataset = \big\{ (x_i, u_i, x_i^+) : i \in \{1,\ldots,N\} \big\},
    \end{equation*}
    where each underlying $\xi_1, \ldots, \xi_N$ is an i.i.d. sample from $(\Omega, \mathcal{F}, \mathbb{P})$.
\end{definition}

Given the dataset $\Dataset$, we consider the following problem.

\begin{problem}[Data-driven synthesis]
\label{problem}
Let $\system$ be a system with a Lipschitz continuous dynamics function $f$ (\cref{def:Lipschitz}), let $x_0 \in \X$ be an initial state, and let $\varphi$ be an scLTL specification.
Given a dataset $\Dataset$ and a confidence parameter $\beta \in (0,1)$, compute a policy $\pi \in \Pi^\system$ and a lower bound $\lambda \in [0,1]$ such that 
\begin{equation*}
  \Prob^N \big\{ (\xi_1, \ldots, \xi_N) \in \Omega^N : 
    \satprob_{\pi,x_0}^\system(\varphi)
     \geq \lambda
    \big\} \geq 1 - \beta.   
\end{equation*}
\end{problem}
\smallskip

\cref{problem} asks for a policy $\pi$ together with a lower bound $\lambda$ on its satisfaction probability.
Since $\lambda = 0$ is a trivial solution, the goal is to find a policy for which $\lambda$ is as large as possible.
The outer probability accounts for the chance that we might (with probability at most $\beta$) use a dataset that leads to an invalid solution, i.e., $\varphi$ is satisfied with probability lower than $\lambda$. 
We note that this confidence probability is unavoidable with any finite-sample method and, regardless, the confidence can be selected to be remarkably close to $1$.

\begin{remark}[Upper bounds]
    Conversely, \tool can also synthesize \emph{upper} bounds on the satisfaction probability.
    For simplicity, we only consider lower bounds throughout this paper.
\end{remark}

\subsection{Data-driven IMDP abstraction}
\tool solves \cref{problem} by learning a finite-state abstraction of the system $\system$ based on the dataset $\Dataset$.
The tool supports two modeling formalisms for the abstraction, both of which are special cases of \emph{robust Markov decision processes} (RMDPs)~\cite{DBLP:journals/ior/WiesemannKS14,suilen_robust_2024}.

\begin{definition}[RMDP]\label{def:RMDP}
    A \emph{robust Markov decision process} (RMDP) $\rmdp$ is a tuple 
    $
    \RMDP,
    $
    where 
    \begin{itemize}
        \item $\States$ is a finite set of states, with $\Sinit \in \States$ the initial state,
        \item $\Actions$ is a finite set of actions,
        \item $\cP \colon \States \times \Actions \to 2^{\distr{\States}}$ is an uncertain transition function with $\cP(s,a) \neq \emptyset$ for all $(s,a) \in \States \times \Actions$, and
        \item $\labelfunc \colon \States \to 2^\Labels$ is a labeling function over the finite set of labels $\Labels$, which is shared with system $\system$.
    \end{itemize}
\end{definition}

Intuitively, an RMDP defines a game between a \emph{strategy} $\policyIMDP$ that chooses the actions $a \in \Actions$, and an \emph{adversary} that chooses the transition probability distributions in $\cP(s,a)$.
Formally, a strategy $\policyIMDP \colon \States^* \to \Actions$ maps finite paths to actions, and an adversary is a function $\tau \colon \States^* \times \Actions \to \distr{\States}$ such that $\tau(s_0 \cdots s_k, a) \in \cP(s_k, a)$ for every finite path $s_0 \cdots s_k \in \States^*$ and action $a \in \Actions$.
Thus, the adversary may choose a different distribution from $\cP(s,a)$ each time the pair $(s,a)$ is visited.
Together, a strategy $\policyIMDP$ and an adversary $\tau$ induce a Markov chain with a unique probability measure $\Prob^{\rmdp}_{\policyIMDP,\tau}$ over the infinite paths $\States^\omega$ starting in $\Sinit$.
As for system $\system$, every path $(s_0, s_1, \ldots)$ corresponds with a unique trace $(\labelfunc(s_0), \labelfunc(s_1), \ldots)$, and the \emph{robust satisfaction probability} of an scLTL specification $\varphi$ under strategy $\policyIMDP$ is the worst case over all adversaries:
\begin{equation}
    \label{eq:RMDP_robust_satprob}
    \satprob^{\rmdp}_{\policyIMDP}(\varphi) \coloneqq \inf_{\tau} \, \Prob^{\rmdp}_{\policyIMDP,\tau}\big\{ \mathbf{y} \in (2^\Labels)^\omega : \mathbf{y} \models \varphi \big\}.
\end{equation}

\begin{remark}[Policies vs. strategies]
    For clarity, we denote policies for the stochastic system $\system$ by $\policy$, while we refer to policies on the abstract models as \emph{strategies} and use the symbol $\policyIMDP$.
\end{remark}

An RMDP is an \emph{interval MDP} (IMDP) if there exist bounds $\Pl, \Pu \colon \States \times \Actions \times \States \to [0,1]$, with $\Pl \leq \Pu$, such that for all $(s,a) \in \States \times \Actions$,
\[
  \mathcal{P}(s,a) = \big\{ \mu \in \distr{\States} :
    \Pl(s,a,s') \leq \mu(s') \leq \Pu(s,a,s') \;\; \forall s' \in \States \big\}.
\]
A \emph{two-layer IMDP} (2IMDP) imposes an additional layer of interval constraints on the transition function.
Let $\mathcal{B} = \{B_1, \ldots, B_{|\mathcal{B}|}\}$, with each $B_i\subseteq \States$, be a partition of $\States$ into \emph{blocks} (\emph{clusters}) and write $\mu(B) := \sum_{s' \in B} \mu(s')$. 
An RMDP is a 2IMDP with respect to $\mathcal{B}$ if there exist state-level bounds $\Pl, \Pu \colon \States \times \Actions \times \States \to [0,1]$ and
block-level bounds $\Pl_{\mathcal{B}}, \Pu_{\mathcal{B}} \colon \States \times \Actions \times \mathcal{B} \to [0,1]$
such that for all $(s,a) \in \States \times \Actions$,
\begin{align*}
  \mathcal{P}(s,a) = \big\{ \mu \in \distr{\States} :\;
    {}&{} \Pl(s,a,s') \leq \mu(s') \leq \Pu(s,a,s') \;\; \forall s' \in \States, \;\;
    \\
    {}&{} \Pl_{\mathcal{B}}(s,a,B) \leq \mu(B) \leq \Pu_{\mathcal{B}}(s,a,B) \;\; \forall B \in \mathcal{B}
  \big\}.
\end{align*}

\begin{figure}[t]
\centering
    \input{figures/tool_overview}
    \caption{Modules within \tool. From a dataset $\Dataset$ generated by a simulator and an scLTL specification $\varphi$, \tool computes a finite-memory policy $\policy$ together with a certified lower bound on the probability that the system satisfies $\varphi$.}
    \label{fig:overview}
\end{figure}
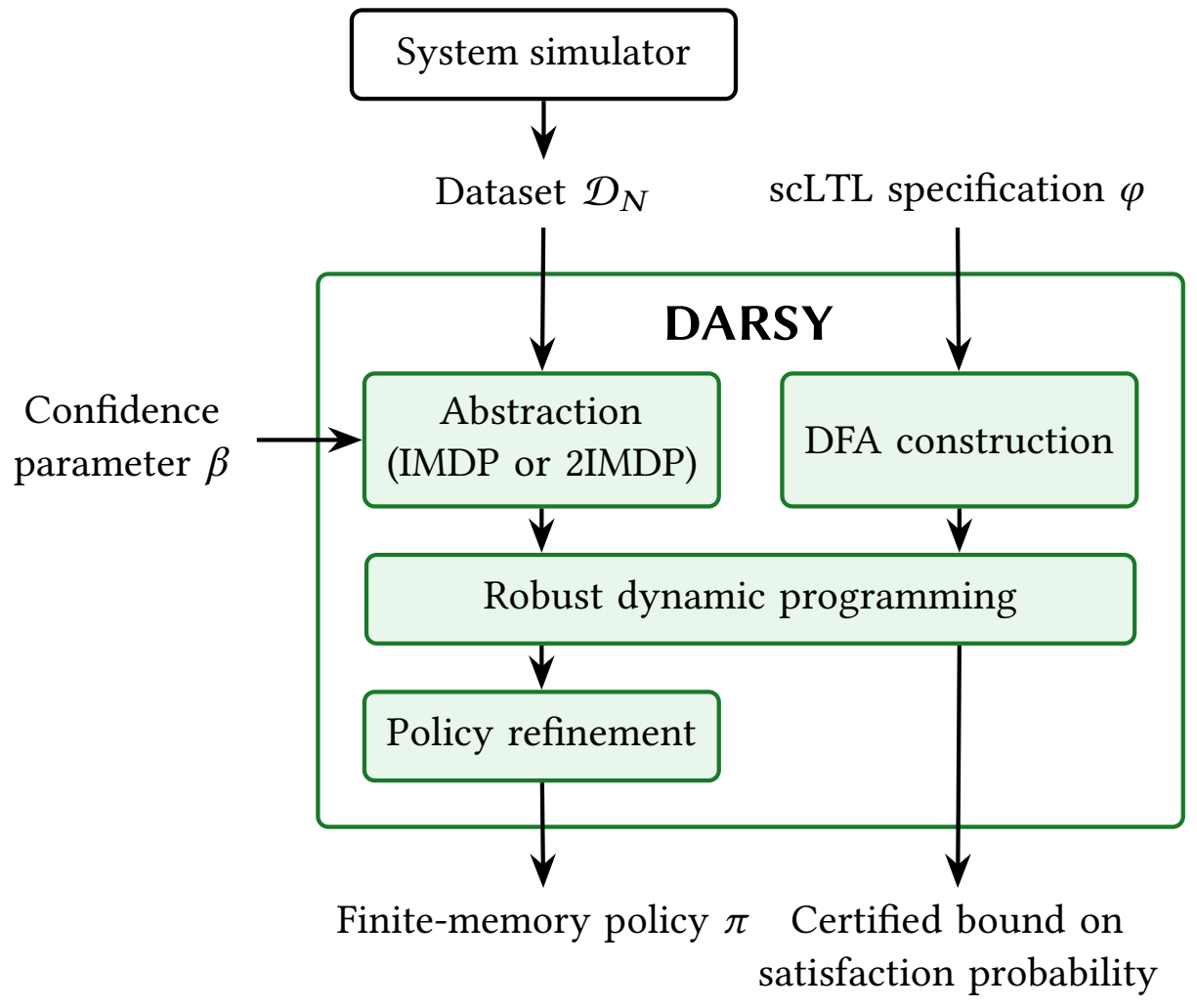

\section{Tool Structure and Algorithms}
\label{sec:algorithms}

\Cref{fig:overview} shows the key modules in \tool.
The tool uses Spot~\cite{DBLP:conf/cav/Duret-LutzRCRAS22} to translate the scLTL specification $\varphi$ into a DFA.
From the dataset~$\Dataset$, \tool then constructs an IMDP or 2IMDP abstraction of the system $\system$ (\cref{sec:abstraction}).
The abstraction algorithm follows~\cite{DBLP:conf/cdc/NazeriBSSA25}, extended with the 2IMDP abstraction and synthesis algorithm from~\cite{DBLP:journals/corr/abs-2507-02213}.
On the product of the abstraction with this DFA, \tool computes an abstract strategy by robust dynamic programming (\cref{sec:synthesis}).
Finally, \tool refines the abstract strategy into a finite-memory policy for $\system$, together with a lower bound on the satisfaction probability that holds with probability at least $1 - \beta$ (\cref{sec:refinement}).
As such, this refined policy solves \cref{problem}.

\subsection{Abstraction into an IMDP or 2IMDP}
\label{sec:abstraction}
\tool constructs the abstraction using the algorithm from~\cite{DBLP:conf/cdc/NazeriBSSA25}.
Since $\X$ may be unbounded, we abstract only a compact hyperrectangular subset $\Xc \subseteq \X$ of the state space, chosen by the user.

\paragraph{States, actions, and labels}
\tool partitions $\Xc$ into a finite set of disjoint and nonempty hyperrectangles $\{R_s\}_{s \in \States \setminus \{s_\star\}}$ (where each boundary point is assigned to one of the adjacent regions), such that the labeling function $g$ is constant on each region. The labeling of the abstraction is thus $\labelfunc(s) = g(x)$ for any $x \in R_s$. By the assumptions on $g$ in \cref{subsec:dtss}, such a partition $\{R_s\}_{s \in \States \setminus \{s_\star\}}$, which we call \emph{$\Y$-respecting}, can always be obtained. An absorbing sink state $s_\star$ with region $R_{s_\star} = \X \setminus \Xc$ and label $\labelfunc(s_\star) = \emptyset$ captures all transitions that leave $\Xc$, and the initial state $\Sinit$ is such that $x_0 \in R_{\Sinit}$.
Since the sink state is treated as violating $\varphi$ (see \cref{sec:synthesis}), all regions labeled by the goal of the specification should lie within $\Xc$.
Since the system $\system$ has a finite input space $\U$, we take these inputs directly as the IMDP actions, i.e., $\Actions = \U$.

\paragraph{Reachable sets from data}
Next, \tool \emph{counts} the number of transitions between abstract states present in the dataset $\Dataset$.
Fix a state $s \neq s_\star$ and action $u$, and let $\Dataset(s,u) \subseteq \Dataset$ be the $M \leq N$ samples $(x_i, u_i, x_i^+)$ with $x_i \in R_s$ and $u_i = u$. %
By \cref{def:Lipschitz}, for every $x \in R_s$, the successor $f(x, u, \noise(\xi_i))$ under the same noise value that generated $x_i^+$ is contained in the ball
\begin{equation*}
    \hat{B}_i = \Big\{ x' \in \RR^n : |x' - x_i^+| \leq L_X \cdot \sup_{\hat{x} \in R_s} |\hat{x} - x_i| \Big\}.
\end{equation*}
Thus, for all $x \in R_s$ simultaneously, the number of samples whose successor from $x$ lies in a region $R_{s'}$ is at least $\check{M} = |\{i : \hat{B}_i \subseteq R_{s'}\}|$ and at most $\hat{M} = |\{i : \hat{B}_i \cap R_{s'} \neq \emptyset\}|$.\footnote{Since $f$ is measurable and each $R_{s'}$ is a hyperrectangle, the events $\hat{B}_i \subseteq R_{s'}$ and $\hat{B}_i \cap R_{s'} \neq \emptyset$ are measurable, so $\check{M}$ and $\hat{M}$ are well-defined random variables.}

\paragraph{Probability intervals}
From these counts, and for a fixed confidence parameter $\beta' \in (0,1)$ (not to be confused with $\beta$ in \cref{problem}), \tool computes the transition probability intervals using the Clopper-Pearson confidence bound~\cite{clopper1934use} as
\begin{subequations}
\label{eq:CP}
\begin{align}
    \Pl(s,u,s') &= \mathrm{Beta}^{-1}\big(\tfrac{\beta'}{2}; \check{M}, M - \check{M} + 1\big),
    \label{eq:CP_lb}
    \\
    \Pu(s,u,s') &= \mathrm{Beta}^{-1}\big(1 - \tfrac{\beta'}{2}; \hat{M} + 1, M - \hat{M}\big),
    \label{eq:CP_ub}
\end{align}
\end{subequations}
where $\mathrm{Beta}^{-1}(\gamma; c_1, c_2)$ is the $\gamma$-quantile of the beta distribution with shape parameters $c_1, c_2$, and with $\Pl = 0$ if $\check{M} = 0$ and $\Pu = 1$ if $\hat{M} = M$.\footnote{We omit the more detailed definition of these probability bounds and instead refer the interested reader to~\cite{DBLP:conf/cdc/NazeriBSSA25} for further details.} 
With probability at least $1 - \beta'$, it holds that 
\[
\Pl(s,u,s') \leq \Prob\{ f(x,u,\noise) \in R_{s'} \} \leq \Pu(s,u,s')
\qquad
\forall x \in R_s.
\]
In other words, if we sample many different datasets $\Dataset$, then for at least a $(1-\beta')$-fraction of them, the obtained probability interval $[\Pl(s,u,s'), \Pu(s,u,s')]$ is correct~\cite{DBLP:conf/cdc/NazeriBSSA25}.

\paragraph{Two-layer IMDP}
To construct a 2IMDP, we group the regions into blocks $\mathcal{B}$, which defines a coarser partition of $\X$ into regions $R_B = \bigcup_{s' \in B} R_{s'}$, with $B\in \mathcal B$. We let $s_\star$ be in a block of its own and apply the same procedure to compute the probability intervals on the target sets $R_B$, yielding the block-level bounds $\Pl_{\mathcal{B}}, \Pu_{\mathcal{B}}$~\cite{DBLP:journals/corr/abs-2507-02213}.
These bounds can be much tighter than the state-level bounds for two reasons.
First, a ball $\hat{B}_i$ intersecting two regions in the same block is not fully contained in either region but can still be fully contained in the block.
Second, since a single interval is computed per block, a block of regions that are rarely reached receives one small upper bound, instead of the sum of the upper bounds of all its regions. The following remark further mitigates this issue.

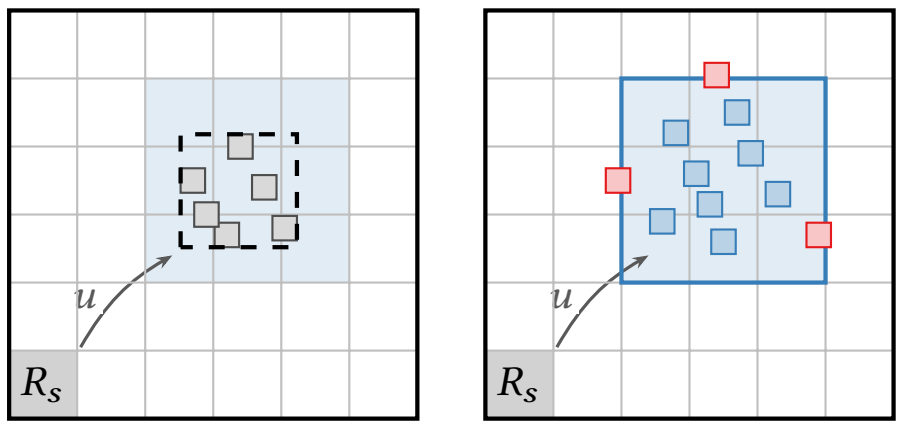
\begin{figure}[t]
    \centering
    \input{figures/two_phase_sampling}
    \caption{Two-phase sampling for a state-action pair $(s,u)$. (a)~The successors of a first batch of samples from $R_s$, each inflated to a box using the Lipschitz constant, define a bounding box (dashed). 
    The regions that intersect this bounding box are the candidate successors $\States_{s,u}$ (in blue). 
    (b)~An independent second batch yields the probability intervals for all $s' \in \States_{s,u}$. 
    The boxes that leave $R_{s,u}$ (red) count towards the probability of transitioning to the sink state $s_\star$.}
    \label{fig:two_phase_sampling}
\end{figure}

\begin{remark}[Two-phase sampling]
    \label{rem:two_step}
    Since the Clopper-Pearson upper bound in \cref{eq:CP_ub} is nonzero even if $\hat{M} = 0$, we would generally obtain a \emph{dense} IMDP, i.e., every $s' \in \States$ can be reached from every state-action pair $(s,a) \in \States \times \Actions$ with nonzero probability. 
    Because robust dynamic programming uses the worst-case transition distributions, this dense abstraction greatly increases conservatism \cite{DBLP:journals/corr/abs-2412-11343} and can lead to poor synthesis outcomes. 
    \tool therefore determines, for each abstract state-action pair $(s,u)$, a set that contains the successor state with high probability and only computes transition probabilities to states in that set. 
    Specifically, \tool leverages \emph{two} independent batches of samples per state-action pair $(s,u)$, as visualized in \cref{fig:two_phase_sampling}.
    The first (small) batch selects \emph{candidate successors} $\States_{s,u} \subseteq \States$ as the regions that intersect the bounding box of the sampled successor states.
    The second batch yields the intervals for all $s' \in \States_{s,u}$, and the probability of leaving $R_{s,u} = \bigcup_{s' \in \States_{s,u}} R_{s'}$ is assigned to the sink $s_\star$, with the upper bound from \cref{eq:CP_ub} for $\hat{M} = |\{i : \hat{B}_i \not\subseteq R_{s,u}\}|$.
    The abstraction remains sound: since the two batches are independent, the Clopper-Pearson bounds are valid for any choice of $\States_{s,u}$, and redirecting transitions to the sink can only decrease the satisfaction probability.
    Moreover, the abstraction is highly \emph{sparse}, and since fewer intervals are computed, all probability bounds are tighter.
\end{remark}

\paragraph{Soundness of the abstraction}
By choosing $\beta' = \beta / n_{\mathrm{int}}$, where $n_{\mathrm{int}}$ is the total number of intervals (including the block-level intervals of a 2IMDP), a union bound ensures that all intervals hold simultaneously with probability at least $1 - \beta$.
In that case, the abstraction is \emph{sound}, i.e., for every $s \neq s_\star$, $u \in \U$, and $x \in R_s$, the distribution over successor regions that assigns $\Prob\{ f(x,u,\noise) \in R_{s'} \}$ to each $s' \in \States_{s,u}$ and the remaining mass $\Prob\{ f(x,u,\noise) \notin R_{s,u} \}$ to the sink $s_\star$ (see \cref{rem:two_step}) is contained in $\cP(s,u)$.
Moreover, every $\cP(s,u)$ is nonempty, as required by \cref{def:RMDP}: since $\Pl \leq \check{M}/M$ and $\Pu \geq \hat{M}/M$ (and similarly for the block-level and sink bounds), the fractions of samples whose successor from any $x \in R_s$ lies in each region $R_{s'}$, $s' \in \States_{s,u}$, or outside $R_{s,u}$ satisfy all constraints.

\subsection{Robust dynamic programming}
\label{sec:synthesis}
\tool uses Spot~\cite{DBLP:conf/cav/Duret-LutzRCRAS22} to translate the scLTL specification $\varphi$ into a DFA $\mathcal{D}_\varphi = (\StatesDFA, q_I, 2^\Y, \PDFA, \StatesDFA_F)$ that accepts its good prefixes, with states $\StatesDFA$, initial state $q_I$, transition function $\PDFA \colon \StatesDFA \times 2^\Y \to \StatesDFA$, and accepting states $\StatesDFA_F \subseteq \StatesDFA$.
The tool then performs strategy synthesis on the product of the abstraction with $\mathcal{D}_\varphi$, whose states are pairs $(s,q) \in \States \times \StatesDFA$.
Under action $a$, the product moves from $(s,q)$ to $(s', \PDFA(q, \labelfunc(s')))$ with probability $\mu(s')$, where $\mu \in \cP(s,a)$ is chosen by an \emph{adversary}.
We call $(s,q)$ \emph{accepting} if $s \neq s_\star$ and $q \in \StatesDFA_F$, so satisfying $\varphi$ amounts to reaching an accepting product state.

\paragraph{Dynamic programming}
\tool computes, via robust dynamic programming, a stationary abstract strategy $\policyIMDP \colon \States \times \StatesDFA \to \Actions$, i.e., its output only depends on the current product $(s,q)$. The returned strategy also maximizes the probability of reaching an accepting product state under the worst-case transition probabilities of the abstraction.\footnote{For infinite horizons, robust dynamic programming converges only in the limit to the optimal values. By conservatively initializing the value function, we ensure that we always approach the optimal values from the conservative direction.}
Let $V^{(0)}(s,q) = 1$ if $(s,q)$ is accepting and zero otherwise.
Accepting product states keep value one, and the states $(s_\star, q)$ keep value zero for all $q \in \StatesDFA$, i.e., leaving $\Xc$ is treated as violating $\varphi$.
This is conservative but sound: traces that satisfy $\varphi$ only after leaving $\Xc$ are not counted, so the computed values can only underestimate the satisfaction probability.
For all other product states $(s,q)$ and for all iterations $j \in \NN$, we consider the recursion
\begin{equation}
    \label{eq:DP}
    V^{(j+1)}(s,q) = \max_{a \in \Actions} \min_{\mu \in \cP(s,a)} \sum_{s' \in \States} \mu(s') \, V^{(j)}\big(s', \PDFA(q, \labelfunc(s'))\big).
\end{equation}
This max-min dynamic programming recursion can be solved efficiently for both IMDPs and 2IMDPs.
The two only differ in the feasible set of distributions in the inner minimization in \cref{eq:DP}:
\begin{itemize}
    \item \emph{IMDP:} The adversary is only constrained by the state-level bounds $\Pl, \Pu$.
    The inner minimization is solved exactly by the standard greedy approach~\cite{DBLP:journals/ai/GivanLD00,suilen_robust_2024}: first assign each successor $s'$ its lower bound $\Pl(s,a,s')$, and then distribute the remaining probability mass over the successors in increasing order of their values, up to their upper bounds.
    \item \emph{2IMDP:} The adversary is additionally constrained by the block-level bounds $\Pl_{\mathcal{B}}(s,a,B) \leq \mu(B) \leq \Pu_{\mathcal{B}}(s,a,B)$.
    In particular, it cannot move more than $\Pu_{\mathcal{B}}(s,a,B)$ of the probability mass into a block $B$, even if the successors in $B$ individually admit more.
    The greedy approach is extended accordingly~\cite{DBLP:journals/corr/abs-2507-02213}: an additional first phase assigns each block the mass needed to meet its lower bound, again in increasing order of value, and in the second phase, the remaining mass is capped by both the state- and block-level upper~bounds.
\end{itemize}
Since the block-level constraints shrink the uncertainty sets, a 2IMDP yields worst-case values that are at least as high as those of an IMDP with the same state-level bounds, i.e., less conservative lower bounds, at a somewhat higher computational cost, since both approaches are dominated by sorting the successors by value.

\paragraph{Termination and strategy extraction}
\tool runs robust dynamic programming until the maximal change between two iterates $V^{(j)}$ and $V^{(j+1)}$ is below a prescribed tolerance.
Upon termination, the tool extracts an optimal strategy $\policyIMDP$ corresponding to the value function $V_\policyIMDP \coloneqq V^{(J)}$ at the final iteration $J$.
Since the value function is initialized conservatively, this final value function $V_\policyIMDP$ soundly lower bounds the worst-case probability that $\policyIMDP$ reaches an accepting product state (modulo floating-point errors).

\subsection{Policy refinement}
\label{sec:refinement}
The abstract strategy $\policyIMDP$ is refined into a finite-memory policy $\pi$ for the system $\system$.
Let $\mathcal{R}(x)$ be the abstract state whose region contains $x$, i.e., $x \in R_{\mathcal{R}(x)}$, so that $\mathcal{R}(x) = s_\star$ for all $x \notin \Xc$, and let $\policyIMDP(s_\star, q)$ be an arbitrary fixed input for all $q \in \StatesDFA$.
The DFA state is tracked along the path as $q_0 = \PDFA(q_I, g(x_0))$ and $q_{k+1} = \PDFA(q_k, g(x_{k+1}))$, and the control input at time $k$ is
\begin{equation*}
    u_k = \pi(x_0, \ldots, x_k) \coloneqq \policyIMDP\big(\mathcal{R}(x_k), q_k\big).
\end{equation*}
Since $\Actions = \U$, the action selected by $\policyIMDP$ is applied directly as the input, and since $g$ is constant on each region, the DFA tracks the same labels on the system as on the abstraction.
\Cref{alg:closed_loop} summarizes the behavior of the resulting closed-loop system.

\begin{algorithm}[t]
\caption{Closed-loop system under the refined policy $\pi$.}
\label{alg:closed_loop}
\begin{algorithmic}[1]
    \Require Abstract strategy $\policyIMDP$, DFA $\mathcal{D}_\varphi$, initial state $x_0$
    \State $q_0 \gets \PDFA(q_I, g(x_0))$ \Comment{Read initial label}
    \For{$k = 0, 1, 2, \ldots$}
        \If{$q_k \in \StatesDFA_F$}
            \State \Return $\varphi$ is satisfied \Comment{Good prefix read}
        \EndIf
        \State $s_k \gets \mathcal{R}(x_k)$ \Comment{$s_k = s_\star$ if $x_k \notin \Xc$}
        \State $u_k \gets \policyIMDP(s_k, q_k)$ \Comment{Retrieve input from policy}
        \State Apply $u_k$ to $\system$ and observe $x_{k+1}$
        \State $q_{k+1} \gets \PDFA\big(q_k, g(x_{k+1})\big)$ \Comment{Update memory}
    \EndFor
\end{algorithmic}
\end{algorithm}

The policy $\pi$ only depends on the current state $x_k$ and the DFA state $q_k$, as discussed in \cref{sec:specs}.
The following result, which combines the union bound from \cref{sec:abstraction} with the abstraction-based guarantees in~\cite{DBLP:conf/cdc/NazeriBSSA25}, shows that $\pi$ and $\lambda = V_\policyIMDP(\Sinit, q_0)$ solve \cref{problem}.

\begin{theorem}[Soundness]
    \label{thm:soundness}
    Let $\pi$ be the refinement of the abstract strategy $\policyIMDP$ computed by \tool, and let $q_0 = \PDFA(q_I, g(x_0))$.
    Then, 
    \begin{equation*}
        \Prob^N \big\{ (\xi_1, \ldots, \xi_N) \in \Omega^N :
            \satprob_{\pi,x_0}^\system(\varphi) \geq V_\policyIMDP(\Sinit, q_0)
        \big\} \geq 1 - \beta.
    \end{equation*}
\end{theorem}

Intuitively, with probability at least $1 - \beta$, all probability intervals of the abstraction are correct, so the abstraction is sound.
The adversary in the abstraction can then mimic the system until it leaves $\Xc$, after which the abstraction assigns value zero, so the robust value $V_\policyIMDP(\Sinit, q_0)$ is a lower bound on the satisfaction probability of the system.
In particular, if $x_0 \notin \Xc$, then $\Sinit = s_\star$ and the theorem holds trivially with $V_\policyIMDP(\Sinit, q_0) = 0$.

\section{Using \tool: Operating Instructions}
\label{sec:usage}
In this section, we describe a typical usage of \tool.
Using a specific benchmark, namely, the \texttt{Pendulum} one, as a running example, we describe the most important features for defining models and specifications, and how to run the algorithms from~\cref{sec:algorithms}.

\subsection{Structure of the codebase}
\Cref{fig:codebase} shows the high-level codebase structure, which has two main entry points: one for IMDP and another for the 2IMDP abstraction.

Each benchmark and specification is contained in a single file in \texttt{benchmarks/}, while the underlying system simulators (which can be shared across benchmarks) are all in \texttt{benchmarks/models.py}.
This structure makes it easy to create different benchmark variants, for example, two different specifications for the same system.

All algorithms live in \texttt{core/}, with the main abstraction algorithms in \texttt{core/abstraction/} and the automaton-to-DFA translation in \texttt{core/automaton/}.
Both the IMDP and 2IMDP entry points run the same pipeline (\texttt{core/pipeline.py}), which builds the abstraction, solves it, and reports the results.
Finally, the codebase contains several validation and plotting functionalities.

\begin{figure}[t]
    \centering
    \begin{minipage}{\columnwidth}
        \small
        \input{figures/codebase}
    \end{minipage}
    \caption{Key files and folders in the codebase of \tool, highlighting the section that describes each module.}
    \label{fig:codebase}
\end{figure}

\subsection{Defining models}
\label{subsec:models}
Taking the \texttt{Pendulum} benchmark~\cite{DBLP:conf/nips/TowersKBCDGKKKP25} as an example, its model is written as a discrete-time stochastic system $\system$, as per~\cref{subsec:dtss}.
At its core, its dynamics are defined by a step function representing the dynamics function $f$. This step function is defined inside the class \texttt{PendulumDynamics}, found in \texttt{benchmarks/models.py}:

\begin{lstlisting}[style=python]
def step(self, state, action, noise):

    new_velo = (1-self.b) * state[1] + \
      (3*self.G / (2*self.l) * jnp.sin(state[0])) * self.tau + \
      (3.0 / (self.m * self.l**2) * action[0]) * self.tau
    new_angle = wrap_theta(state[0] + self.tau * new_velo + \
      noise[0])
    new_velo = new_velo + noise[1]

    return jnp.array([new_angle, new_velo])
\end{lstlisting}

The noise distribution is defined in \texttt{benchmarks/Pendulum.py} and is by default a triangular distribution with halfwidths $[0.03,0.1]$:
\begin{lstlisting}[style=python]
self.noise = TriangularDistr(np.array([0.03, 0.1]))
\end{lstlisting}
This stochastic noise and its properties are specified as the class \texttt{TriangularDistr}, which can be found inside
the file \newline \texttt{benchmarks/dynamics/distributions.py}. 
Currently, \tool supports Gaussian, truncated Gaussian, and triangular distributions, but can be extended to support other noise distributions.

The subset $\Xc$ of the state space to partition is specified as the attribute \texttt{partition['boundary']} of the class \texttt{Pendulum} in \texttt{benchmarks/Pendulum.py}. For example, the following line sets $\Xc = [-\pi,\pi]\times[-8,8]$, which stands for an angle component in the range $[-\pi,\pi]$ and an angular velocity in $[-8,8]$:
\begin{lstlisting}[style=python]
self.partition['boundary'] = np.array([[-np.pi, -8], [np.pi, 8]])
\end{lstlisting}
Currently, \tool only supports hyperrectangular $\Xc$. 

The state-space of \texttt{Pendulum} is periodic in its angle (the first state dimension), which is specified in the \texttt{PendulumDynamics} class in \texttt{benchmarks/models.py} by setting the state wrapping:
\begin{lstlisting}[style=python]
self.wrap = jnp.array([True, False], dtype=bool)
\end{lstlisting}
While our problem formulation in \cref{sec:theory} is defined for a finite input set $\U$, our implementation actually supports hyperrectangular input sets $\U$, which are discretized via a uniform gridding. In the case of the \texttt{Pendulum} class, this hyperrectangle as well as the size of its discretization are set as the following attributes:
\begin{lstlisting}[style=python]
self.uMin = [-2]
self.uMax = [2]
self.num_actions = [3]
\end{lstlisting}

\begin{remark}[Unknown dynamics]
Although the step function above contains the system dynamics, the data-driven abstraction algorithms in \tool to solve \cref{problem} use only sampled transitions from these dynamics, together with a precomputed Lipschitz constant $L_X$ of the dynamics. For our \texttt{Pendulum} benchmark, the latter is stored as the following array attribute of the \texttt{PendulumDynamics} class found in \texttt{benchmarks/models.py}:
\begin{lstlisting}[style=python]
self.lipschitz = np.array([[1.15, 0.1],[1.5, 1.0]])
\end{lstlisting}
\end{remark}

\paragraph{Abstraction settings}
For the abstraction, we set a desired resolution for the state-space partition used for abstraction.
For example, the \texttt{Pendulum} benchmark uses a partition of $200 \times 200$ cells, which is specified as an attribute of the \texttt{Pendulum} class:
\begin{lstlisting}[style=python]
self.partition['number_per_dim'] = np.array([200, 200])
\end{lstlisting}

When generating a 2IMDP abstraction, we additionally set the option \cmd{--coarse_factor} to the number of grid cells (per dimension) contained in the same block. 
For example, \cmd{--coarse_factor 3} means every block contains $3^n$ grid cells (with $n$ the state dimension), so the blocks together again define a partition.

\subsection{Defining specifications}
\label{subsec:specifications}
Specifications are defined by the label set $\Y$ and the labeling function $g \colon \X \to 2^{\Y}$.
Both are stored as dictionaries and later translated into a DFA using Spot.
For example, the reach-avoid specification for the \texttt{Pendulum} benchmark has the atomic propositions \texttt{"goal"} and \texttt{"unsafe"}, which are set as attributes of the \texttt{Pendulum} class:
\begin{lstlisting}[style=python]
self.atomic_propositions = ["goal", "unsafe"]
\end{lstlisting}
For each label, we define the subset of the state space where it holds as a set of hyperrectangles, also inside the \texttt{Pendulum} class. 
For example, the following attributes assign the labels $\{\mathsf{goal}\}$ and $\{\mathsf{unsafe}\}$ to the regions $\mathsf{r\_goal}$ and $\mathsf{r\_unsafe}$, respectively, and define $\mathsf{r\_goal}$ as the single rectangle $[-0.1\pi,0.1\pi]\times[-1,1]$ and $\mathsf{r\_unsafe}$ as an empty set:

\begin{lstlisting}[style=python]
self.labels = {
        "r_goal": {"goal"},
        "r_unsafe": {"unsafe"}
    }

self.regions = {
        "r_goal": np.array([
        [[-0.1*np.pi, -1], [0.1*np.pi, 1]]
        ], dtype=float),

        "r_unsafe": np.array([
        ], dtype=float)
    }
\end{lstlisting}

Finally, the scLTL formula, which for \texttt{Pendulum} is reach-avoid, is also set in the \texttt{Pendulum} class:

\begin{lstlisting}[style=python]
self.specification = ("(!unsafe) U (goal)")
\end{lstlisting}

Thanks to the integration of Spot to translate specifications into DFAs, \tool supports general scLTL formulas.

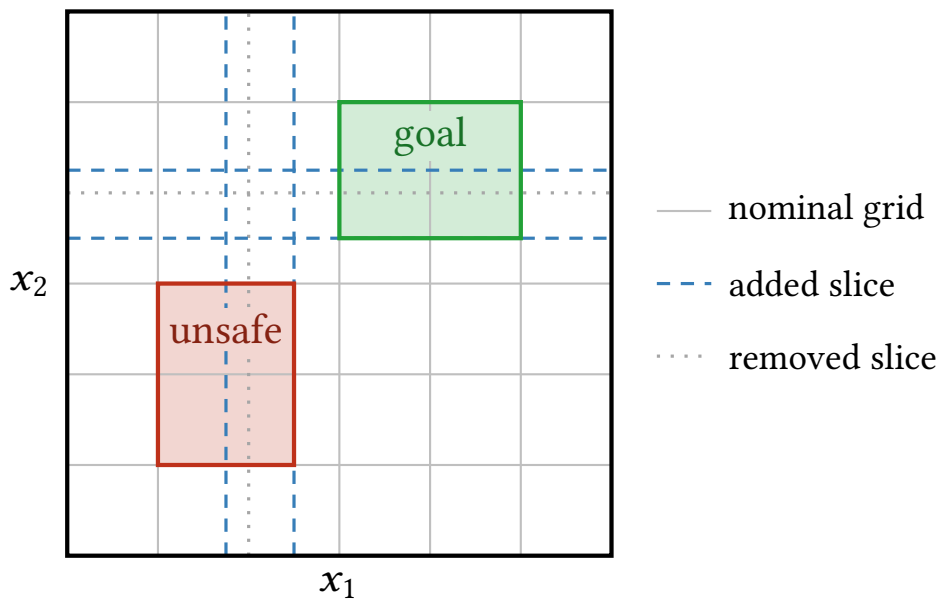
\begin{figure}[t]
\centering
    \input{figures/partition}
    \caption{An example (non-uniform) state partition generated by \tool.
    The user-requested $6 \times 6$ grid is not aligned with the right face of the unsafe region and the bottom face of the goal region. 
    Thus, the tool creates additional slices through these faces and normalizes the cells within each region of interest (shown as dashed blue lines) so that all cells in the same region of interest (goal and unsafe) have the same size.}
    \label{fig:partition}
\end{figure}

\paragraph{Non-uniform state partitions}
To ensure soundness of the abstraction, all states $x,x'$ within the same partition element must have the same labels. 
For example, in the case of \texttt{Pendulum}, all states $x\in \mathsf{r\_goal}$ must satisfy $g(x) = \{\mathsf{goal}\}$.
However, the grid requested by the user may not align with this label-preserving requirement.
To this end, the tool adapts the user-requested grid into a non-uniform grid, as illustrated in \cref{fig:partition}.
Specifically, the faces of all labeled regions (which are hyperrectangles) become grid lines, and the intervals between consecutive grid lines are split uniformly so that no cell is wider than in a user-specified nominal grid.
By construction, this procedure yields a non-uniform grid that is label-preserving and thus suited for abstraction.

\subsection{Synthesizing abstractions and strategies}
Constructing the abstraction and synthesizing an abstract strategy for a desired benchmark can be done by running a single Python file with the desired arguments.
For example, a minimal IMDP run for \texttt{Pendulum} with the default arguments amounts to:
\begin{lstlisting}[style=python]
python RunIMDP.py --model Pendulum
\end{lstlisting}
Similarly, a minimal run for the equivalent 2IMDP abstraction with default arguments amounts to:
\begin{lstlisting}[style=python]
python Run2IMDP.py --model Pendulum
\end{lstlisting}

To avoid materializing the full dataset before constructing the abstraction, \tool generates the $M$ samples $(x_i, u_i, x_i^+)$ in $\Dataset(s,u)$ for every state-action pair $(s,u)$ separately within the abstraction construction loop.
From these samples, the tool constructs the abstract IMDP (or 2IMDP) and synthesizes an optimal abstract strategy, using the techniques from~\cref{sec:abstraction,sec:synthesis}, respectively.

\begin{remark}
    While the current tool supports datasets $\Dataset$ obtained by sampling from a system simulator, our techniques can readily be used on a fixed dataset, for example, obtained from experimental system data (as long as the data forms a valid $\Dataset$ as per \cref{def:dataset}).
\end{remark}

\paragraph{Dedicated solver paths}
For two common classes of specifications, namely \emph{safety} and \emph{reach-(while-)avoid}, the general product construction with a DFA is unnecessary.
For these specifications, \tool bypasses DFA construction and calls a dedicated solver instead, which reduces both memory usage and runtime.
\begin{itemize}
    \item \textbf{Safety:}
    A finite-horizon safety specification is defined by the formula \texttt{G(safe)},\footnote{We remark that this is not scLTL notation, but use it for the sake of conciseness.} a horizon $\horizon$, and a union of safe hyperrectangles.
    A safety specification is interpreted as $\bigwedge_{k=0}^{\horizon} \Next^k \mathsf{safe}$ (see \cref{sec:specs}) and solved by $\horizon$ steps of robust dynamic programming. 
    For instance, the \texttt{Temperature2D} class defines the specification and horizon~as
\begin{lstlisting}[style=python]
self.specification = "G(safe)"
self.horizon = 15
\end{lstlisting}
    Synthesis yields a time-varying strategy $\policyIMDP \colon \{1, \ldots, \horizon\} \times \States \to \Actions$, in which the number of remaining steps replaces the role of the DFA state, such that the policy refinement in \cref{sec:refinement} and \cref{thm:soundness} apply analogously.
    \item \textbf{Reach-avoid:}
    If the provided specification is the reach-avoid formula \texttt{(!unsafe) U (goal)}, with the labels (exactly) named \texttt{unsafe} and \texttt{goal}, then \tool solves the recursion in \cref{eq:DP} directly on the abstraction, i.e., without the DFA component. The solver assigns values of one and zero to goal and unsafe (including $s_\star$) states, respectively. The resulting strategy is memoryless, i.e., $\policyIMDP \colon \States \to \Actions$.
\end{itemize}
Similarly, \tool can compute an upper bound on the satisfaction probability of the synthesized strategy by evaluating it against the most favorable distributions in $\cP(s,a)$.
All other scLTL specifications are handled by the DFA-based procedure from \cref{sec:synthesis}.

\paragraph{Options}
Various options can be passed to the \texttt{RunIMDP.py} and \texttt{Run2IMDP.py} entry points to choose the benchmark to run, the confidence parameter $\beta$, the number of samples per state-action pair used to compute probability intervals, and so on.
A full overview of these options is provided in the documentation.

\subsection{Simulating synthesized policies}
\tool can simulate a system under a synthesized policy and plot results of interest.
The tool contains a Monte Carlo simulator that runs \cref{alg:closed_loop} a specified number of times.
This functionality lets the user obtain an empirical estimate of the probability of satisfying the specification under the synthesized policy, which is helpful for the validation of the certified lower bounds on the satisfaction probability.
Furthermore, the tool includes general plotting functions for heatmaps (e.g., satisfaction probabilities over the state space) and state trajectory plots (e.g., a set of Monte Carlo simulations), as well as tailored plotting functions for specific benchmarks, such as animating a \texttt{Pendulum} trajectory.

\begin{table}[t]
    \centering
    \caption{Benchmark overview ($n$ is the state space dimension).}%
    \label{tab:benchmarks}
    \setlength{\tabcolsep}{3pt}%
    \resizebox{\linewidth}{!}{%
        \begin{tabular}{@{}lcllll@{}}
        \toprule
        \textbf{System} & $n$ & \textbf{Dynamics} & \textbf{Noise} & \textbf{Distribution} & \textbf{Origin} \\
        \midrule
        \textbf{Temp1D}      & 1 & Linear    & Additive       & Gaussian         & \cite{DBLP:journals/corr/abs-2604-13377} \\ 
        \textbf{Temp2D}      & 2 & Linear    & Multiplicative & Trunc.\ Gaussian & \cite{DBLP:journals/corr/abs-2507-02213} \\
        \textbf{Temp3D}      & 3 & Linear    & Multiplicative & Trunc.\ Gaussian & \cite{DBLP:journals/corr/abs-2507-02213} \\
        \textbf{Temp4D}      & 4 & Linear    & Multiplicative & Trunc.\ Gaussian & \cite{DBLP:journals/corr/abs-2507-02213} \\
        \textbf{DoubleInt}   & 2 & Linear    & Additive       & Triangular       & \cite{DBLP:journals/jair/BadingsRAPPSJ23} \\ 
        \textbf{MountainCar} & 2 & Nonlinear & Additive       & Triangular       & \cite{DBLP:conf/nips/TowersKBCDGKKKP25} \\
        \textbf{Pendulum-T}    & 2 & Nonlinear & Additive       & Triangular       & \cite{DBLP:conf/nips/TowersKBCDGKKKP25} \\
        \textbf{Pendulum-G}    & 2 & Nonlinear & Additive       & Gaussian       & \cite{DBLP:conf/nips/TowersKBCDGKKKP25} \\
        \textbf{Dubins2D}    & 2 & Nonlinear & Additive       & Triangular       & \cite{DBLP:journals/corr/abs-2412-11343} \\
        \textbf{Dubins3D}    & 3 & Nonlinear & Additive       & Triangular       & \cite{DBLP:conf/cdc/BadingsA25} \\
        \bottomrule
        \end{tabular}%
    }
\end{table}

\section{Evaluation of \tool}
\label{sec:experiments}

We perform numerical experiments on a variety of benchmarks (see~\cref{tab:benchmarks}) to answer the following questions about \tool:
\begin{enumerate}[label=\textbf{Q\arabic*:}]
    \item Can \tool effectively synthesize policies for systems with unknown dynamics from sampled transitions?
    \item How does the number of samples used to estimate each abstract probability interval affect the results?
    \item Can richer abstractions such as 2IMDPs produce tighter bounds on satisfaction probabilities than conventional IMDPs?
\end{enumerate}

\paragraph{Reproducibility}
All experiments ran with Python 3.13.13 and JAX 0.12.0.1 on an Apple MacBook Pro with an M5 chip and \SI{24}{\giga\byte} of RAM.
The source code is available at \url{https://github.com/tbadings/darsy}, including a script to run all benchmarks and generate \cref{tab:main}.

\paragraph{Benchmarks}
We consider the 10 systems from \cref{tab:benchmarks}, covering a broad range of dynamics classes (linear and nonlinear, additive and multiplicative stochastic noise, and different noise distributions).
We consider different scLTL, reach-avoid, and safety specifications, yielding 13 benchmarks in total.
For each benchmark, we run \tool for an IMDP and a 2IMDP abstraction, with $1\,000$, $10\,000$, and $100\,000$ sampled transitions per abstract state-action pair.
For example, $1\,000$ samples means that, for every pair $(s,a)$, we generate $1\,000$ triples $(x,u,x^+)$, where $x$ is the center of the cell $R_s$ of abstract state $s$ and $u = a$.
We then compute the probability intervals for all $(s,a,s')$ from these $1\,000$ samples.
We run all benchmarks with $\beta=0.001$, i.e., each abstraction, and thus the satisfaction probability, is correct with confidence at least $99.9\%$.

\begin{table*}[t]
    \centering
    \caption{Results for different numbers of sampled transitions per abstract state-action pair, showing the number of abstract states ($|\States|$) and actions ($|\Actions|)$, the number of transitions $(s,a,s')$ in the abstraction (Trans.), the wall-clock time for the abstraction (Abs.) and abstract policy synthesis (Syn.), the lower bound satisfaction probability ($\lambda$), and its empirical estimate (Emp.).}
    \label{tab:main}
    \setlength{\tabcolsep}{3.5pt}
    \ra{0.88}%
    \resizebox{\textwidth}{!}{\input{tables/table_main_extended}}
\end{table*}

\paragraph{Overview of results}
\cref{tab:main} shows the results for all benchmarks, including abstraction sizes, abstraction and synthesis runtimes, lower bounds on satisfaction probability, and the empirical satisfaction over $1\,000$ Monte Carlo simulations.
Although sampling from the system simulator is random, we observed negligible differences across random seeds and therefore report results for a single seed.

\begin{figure}[t]
    \centering
    \begin{subfigure}[t]{0.48\linewidth}
        \centering
        \makebox[\linewidth][l]{\hspace{0.1442\linewidth}\includegraphics[width=0.8349\linewidth]{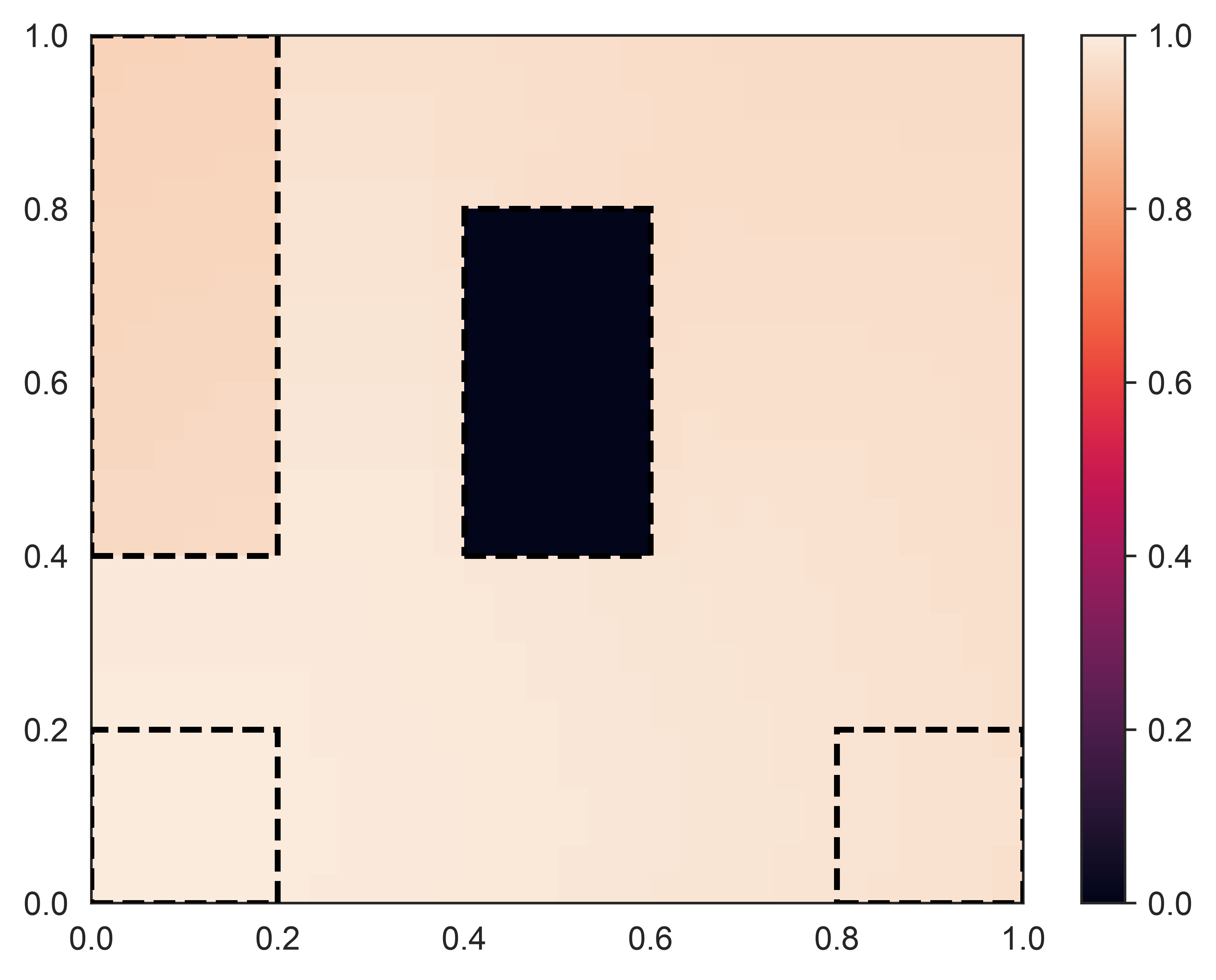}}\\[2pt]
        \makebox[\linewidth][l]{\hspace{0.0209\linewidth}\includegraphics[width=0.8296\linewidth]{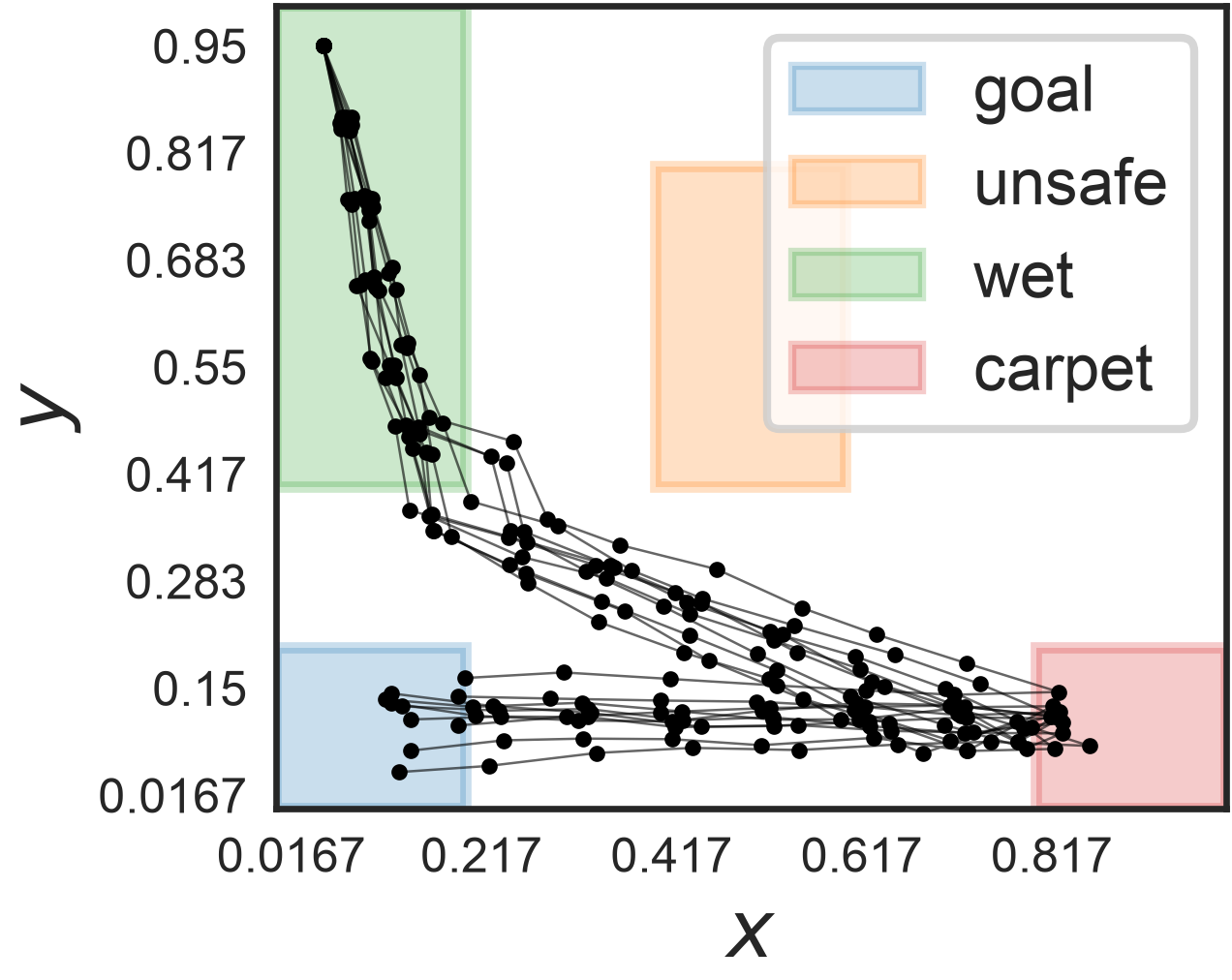}}\par
        \vspace{0.0743\linewidth}%
        \caption{Dubins2D}
        \label{fig:results:dubins}
    \end{subfigure}\hfill
    \begin{subfigure}[t]{0.48\linewidth}
        \centering
        \makebox[\linewidth][l]{\hspace{0.1398\linewidth}\includegraphics[width=0.8602\linewidth]{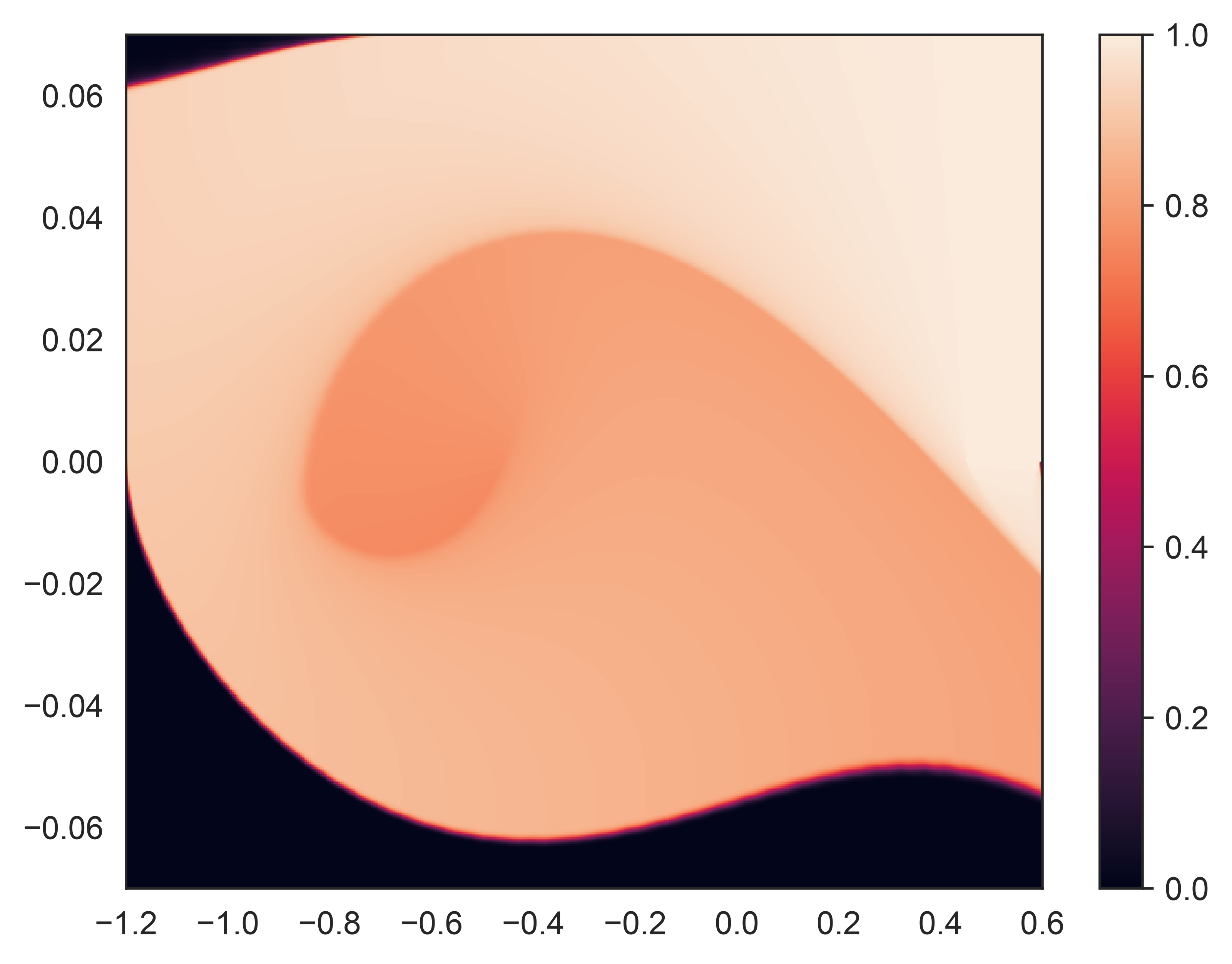}}\\[2pt]
        \makebox[\linewidth][l]{\hspace{0.0000\linewidth}\includegraphics[width=0.9165\linewidth]{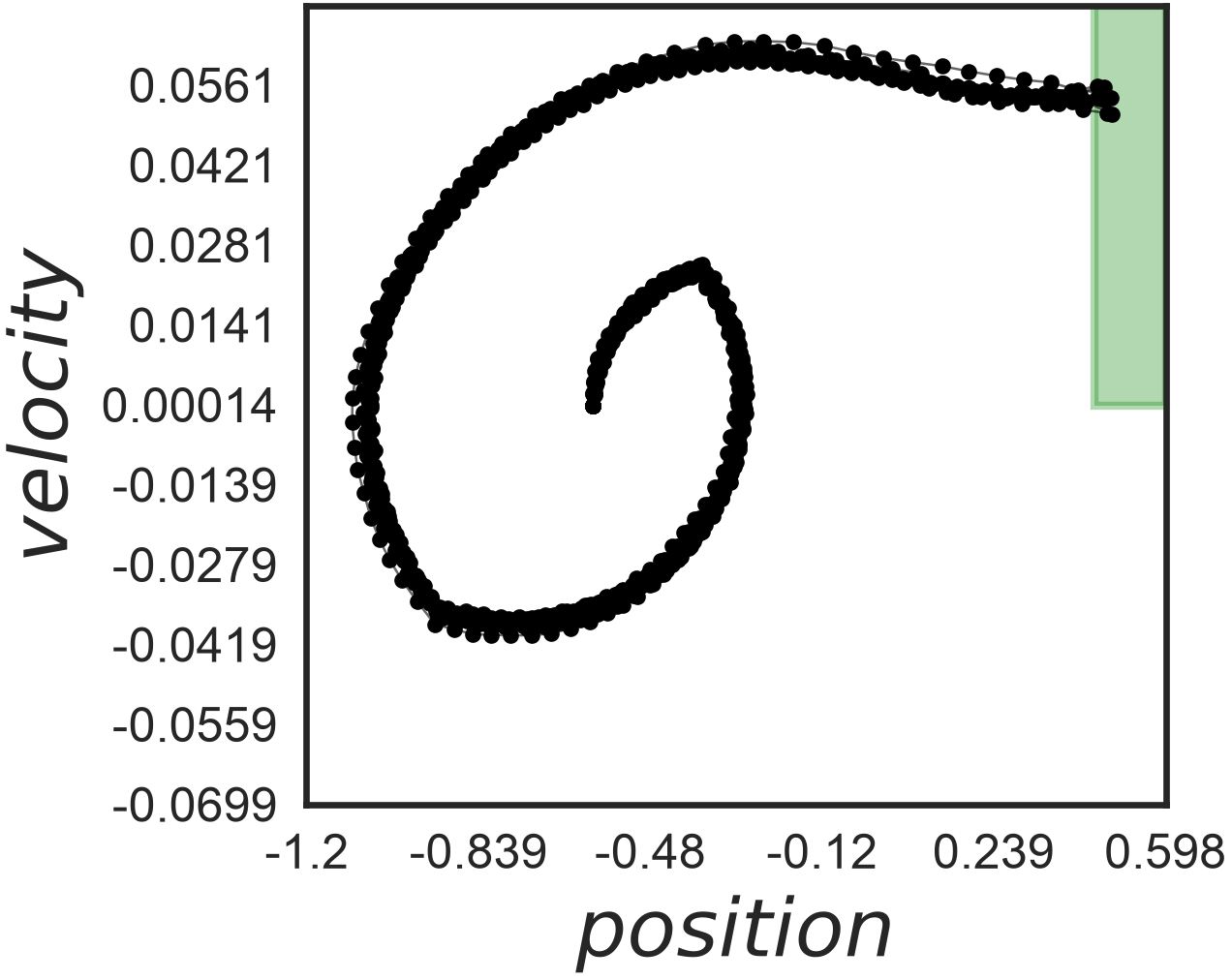}}
        \caption{MountainCar}
        \label{fig:results:mountaincar}
    \end{subfigure}
    \caption{Lower bounds on the satisfaction probability (top) and simulated trajectories (bottom) for two benchmarks, using IMDP abstractions with $10$k samples per state-action pair.}
    \label{fig:results}
\end{figure}

\boldparagraph{Q1: Synthesis performance}
\tool successfully generates abstractions and synthesizes policies for all benchmarks, and all bounds $\lambda$ indeed lower-bound the empirical estimates.
The abstract models have up to roughly $250$k states and $436$ million transitions (i.e., edges in the underlying graph).
\cref{fig:results} shows the lower bound on the satisfaction probability (as a heatmap over initial states $x_0\in \Xc$) together with simulated closed-loop trajectories.
For Dubins2D, the scLTL specification requires reaching $\mathsf{goal}$ while avoiding $\mathsf{unsafe}$, with the additional constraint that if $\mathsf{wet}$ is entered, then $\mathsf{carpet}$ must be visited before reaching $\mathsf{goal}$ \cite{vazquez2018learning}.
Since the initial state lies in $\mathsf{wet}$, the policy indeed steers the system to $\mathsf{carpet}$ first.
MountainCar, in contrast, has a plain reach-avoid specification (reaching the top of the hill~\cite{DBLP:conf/nips/TowersKBCDGKKKP25}).
Its main difficulty is that reaching the goal takes many steps, thus requiring a fine-grained discretization and many samples to obtain a tight bound on the satisfaction probability (with the IMDP, $\lambda = 0.770$ for 10k samples, improving to $\lambda = 0.974$ for 100k). 
The benchmark DoubleInt is also run with an scLTL representing a sequential reachability task.

\begin{figure}[t]
    \centering
    \begin{subfigure}[t]{0.45\linewidth}
        \centering
        \includegraphics[width=\linewidth]{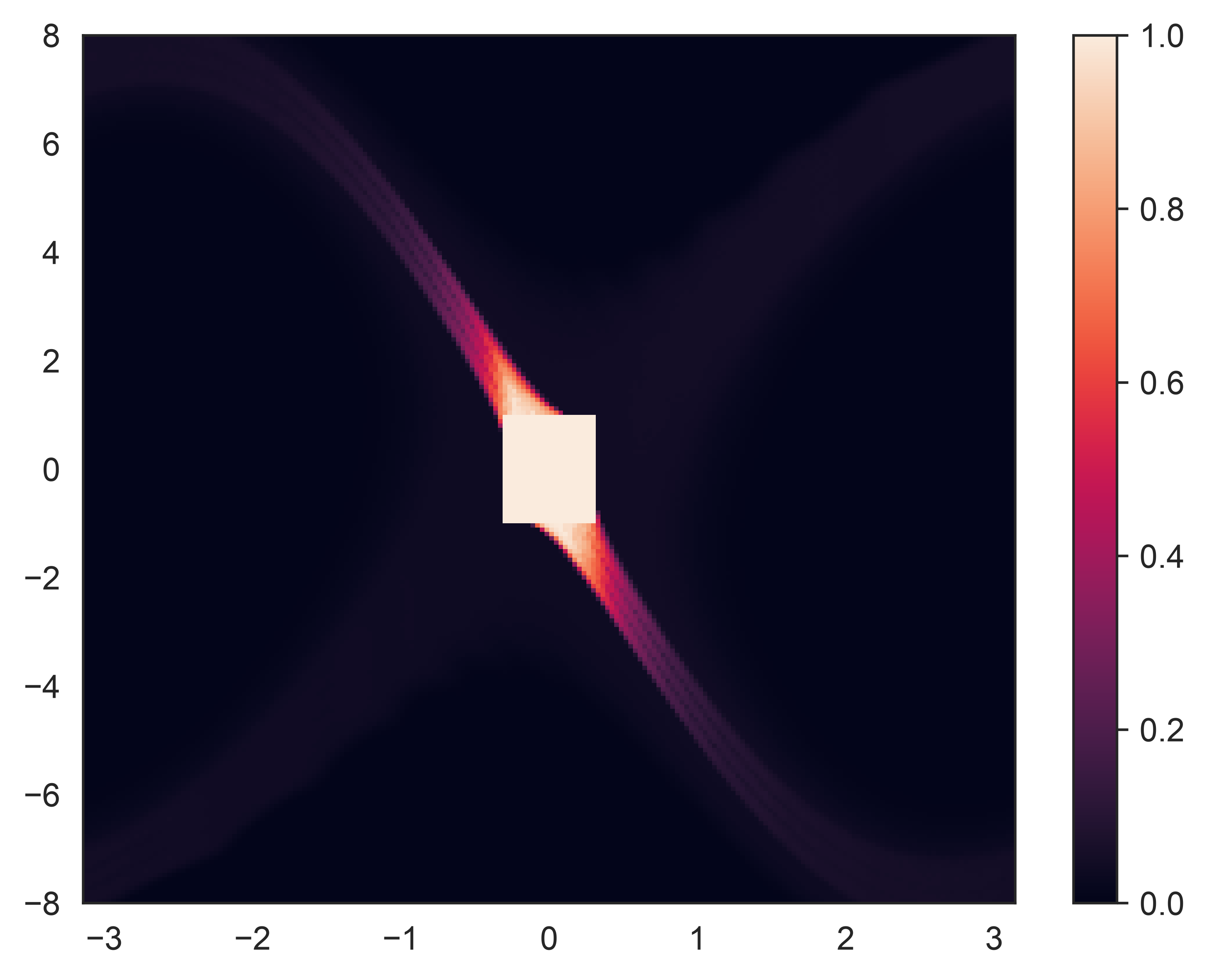}
        \caption{IMDP}
        \label{fig:pendulumhard:imdp}
    \end{subfigure}\hfill
    \begin{subfigure}[t]{0.45\linewidth}
        \centering
        \includegraphics[width=\linewidth]{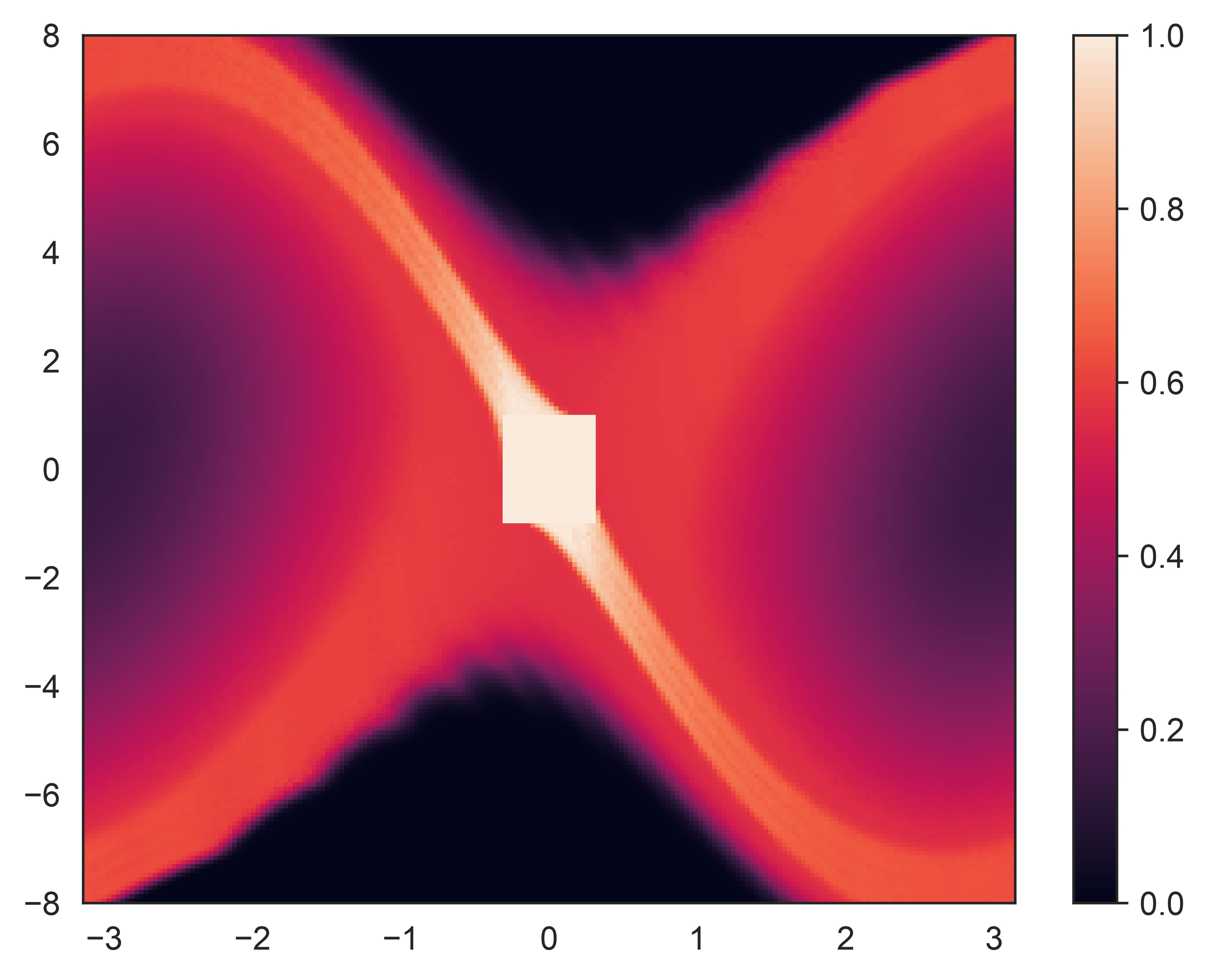}
        \caption{2IMDP}
        \label{fig:pendulumhard:2imdp}
    \end{subfigure}
    \caption{Lower bounds on the satisfaction probability for Pendulum-G (reach-avoid), using an IMDP (left) and a 2IMDP (right) abstraction ($10$k samples per state-action pair).}
    \label{fig:pendulumhard}
\end{figure}

\boldparagraph{Q2: Role of number of sampled transitions}
The lower bound $\lambda$ on the satisfaction probability increases strongly with the number of sampled transitions.
This is as expected, because the tightness of the probability intervals obtained from the Clopper-Pearson confidence bound (see~\cref{eq:CP}) is directly determined by the number of samples.
At the same time, the time to generate abstraction also increases with the number of samples, roughly proportionally, since sampling transitions and counting them is the dominant factor in the abstraction construction.
One exception is Temperature4D, where the 2IMDP abstraction time increases from $171.3$ ($10$k samples) to $4988.3$ seconds ($100$k samples); an increase of almost 30 times.
The reason for this larger increase is likely that this benchmark required more memory than the available \SI{24}{\giga\byte} of RAM, so the machine additionally used the SSD as (slower) swap memory.
Overall, each benchmark shows a sweet spot for the number of sampled transitions, where the bound on the satisfaction probability saturates.
Beyond this point, the gain in satisfaction probability is marginal while abstraction time continues to increase.

\boldparagraph{Q3: IMDP vs 2IMDP abstractions}
We explore the conservatism of 2IMDP abstractions w.r.t. IMDPs.
From \cref{tab:main}, we observe that only Temperature1D and Pendulum-G show a much better bound for 2IMDPs, whereas the improvement for most other benchmarks is smaller, typically a few percent or less.
In some cases, the 2IMDP even shows a marginally smaller satisfaction probability, e.g., for Temperature2D, which is due to the union bound being taken over both state and block-level bounds (thus yielding slightly weaker probability intervals for the same overall confidence).
Nevertheless, the IMDP abstractions for Temperature1D and Pendulum-G yield poor bounds, whereas the 2IMDP bounds are significantly better.
For example, an IMDP abstraction of Temperature1D (reach-avoid) with $100\,000$ samples yields $\lambda = 0.000$, whereas the 2IMDP abstraction gives a much better bound of $\lambda = 0.738$.
Similarly, Pendulum-G with $100$k samples improves from $\lambda = 0.622$ (IMDP) to $\lambda = 0.743$ (2IMDP).
The improvement for Pendulum-G with $10$k samples is shown in~\cref{fig:pendulumhard}.
These results show the benefits of 2IMDPs for obtaining much tighter bounds on satisfaction probabilities on benchmarks that are too challenging for IMDP abstractions.

\section{Conclusion}
We present \tool, a new Python tool for verification and policy synthesis for stochastic systems via data-driven abstractions.
Unlike existing abstraction tools, \tool does not require knowledge of the system dynamics and instead constructs abstractions from sampled transitions.
\tool is automated and sound, providing probably approximately correct guarantees on synthesized policies.
Our experiments on a variety of benchmarks and specifications have shown the tool's generality and broad applicability.

In the future, we plan to extend \tool to gray-box settings, where structural assumptions and partial knowledge of the dynamics can improve performance and data efficiency.
Furthermore, since our JAX implementation vectorizes both the abstraction and dynamic programming, we plan to evaluate \tool on GPUs to scale to higher-dimensional systems and finer~partitions.

\bibliographystyle{ACM-Reference-Format}
\bibliography{references}

\end{document}

%% file: include/packages.tex
\usepackage{nicefrac}
\usepackage{mleftright}
\usepackage{cases}
\usepackage{empheq}
\usepackage{stmaryrd}
\usepackage{bm}
\usepackage{bbm}
\usepackage{mathtools}
\usepackage{accents}
\usepackage[colorinlistoftodos]{todonotes}
\usepackage{dsfont}

\usepackage{siunitx}
\usepackage{subcaption}

\usepackage{xspace}

\usepackage{algorithm}
\usepackage{algpseudocode}
\algrenewcommand\algorithmicrequire{\textbf{Input:}}

\usepackage{etoolbox}
\algtext*{EndFor}
\algtext*{EndIf}
\algtext*{EndWhile}
\algtext*{EndLoop}
\algtext*{EndFunction}
\algtext*{EndProcedure}
\definecolor{algrule}{gray}{0.65}
\makeatletter
\newcommand*{\ALG@blockrule}[1]{%
  \makebox[#1][l]{\hspace*{.35em}\raisebox{0pt}[0pt][0pt]{\color{algrule}\vrule width .4pt height .75\baselineskip depth .3\baselineskip}}}
\newcount\ALG@ruleidx
\newdimen\ALG@enddepth
\def\ALG@printindent{%
  \ifx\ALG@text\ALG@x@notext
    \vrule width\z@ height\z@ depth\ALG@enddepth
  \else
    \ifnum\theALG@nested>0
      \ALG@ruleidx=1
      \loop
        \ALG@blockrule{\csname ALG@ind@\the\ALG@ruleidx\endcsname}%
        \advance\ALG@ruleidx 1
      \ifnum\ALG@ruleidx<\numexpr\theALG@nested+1\relax
      \repeat
    \fi
  \fi}
\patchcmd{\ALG@doentity}{\noindent\hskip\ALG@tlm}{\noindent\ALG@printindent}{}%
  {\PackageWarning{algorule}{Could not patch \string\ALG@doentity}}
\patchcmd{\ALG@doentity}{\item[]\nointerlineskip}%
  {\item[]\ALG@enddepth\prevdepth
   \ifdim\ALG@enddepth>\z@\vskip-\ALG@enddepth\else\ALG@enddepth\z@\fi
   \nointerlineskip}{}%
  {\PackageWarning{algorule}{Could not patch \string\ALG@doentity\space (notext)}}
\makeatother

\usepackage{cleveref}

\crefname{equation}{Eq.}{Eqs.}
\crefname{pluralequation}{Eqs.}{Eqs.}

\crefname{algorithm}{Algorithm}{Algorithm}

\crefname{figure}{Fig.}{Figs.}
\crefname{pluralfigure}{Figs.}{Figs.}

\crefname{section}{Sect.}{Sects.}
\crefname{pluralsection}{Sects.}{Sects.}

\crefname{table}{Table}{Table}
\crefname{pluraltable}{Tables}{Tables}

\crefname{definition}{Def.}{Def.}
\crefname{pluraldefinition}{Defs.}{Defs.}

\crefname{theorem}{Theorem}{Theorems}
\crefname{pluraltheorem}{Theorems}{Theorems}

\crefname{lemma}{Lemma}{Lemmas}
\crefname{plurallemma}{Lemmas}{Lemmas}

\crefname{example}{Example}{Example}
\crefname{pluralexample}{Examples}{Examples}

\crefname{problem}{Problem}{Problem}
\crefname{pluralproblem}{Problems}{Problems}

\crefname{assumption}{Assumption}{Assumption}
\crefname{pluralassumption}{Assumptions}{Assumptions}

\crefname{remark}{Remark}{Remark}
\crefname{pluralremark}{Remarks}{Remarks}

\crefname{proposition}{Proposition}{Proposition}
\crefname{pluralproposition}{Propositions}{Propositions}

\crefname{appendix}{Appendix}{Appendices}
\crefname{pluralappendix}{Appendices}{Appendices}

\usepackage{tikz} 
\usepackage{pgfplots}
\usepgfplotslibrary{external,fillbetween,colorbrewer}
\pgfplotsset{compat=newest}

\DeclareUnicodeCharacter{2212}{−}
\usepgfplotslibrary{external,}
\usetikzlibrary{patterns,shapes.arrows}

\newcolumntype{H}{>{\setbox0=\hbox\bgroup}c<{\egroup}@{}}
\usepackage{colortbl}
\definecolor{LightCyan}{rgb}{0.88,1,1}

\definecolor{red}{rgb}{0.745,0.192,0.102}
\definecolor{darkgreen}{RGB}{34,161,55}
\definecolor{ruhuisstijlrood}{rgb}{0.745,0.192,0.102}
\definecolor{ruhuisstijlzwart}{rgb}{0,0,0}
\definecolor{ruhuisstijlwit}{rgb}{0.98,0.98,0.98}

\usepackage{mdframed}

\usepackage{listings}
\definecolor{codebg}{RGB}{248,248,248}
\definecolor{codeframe}{RGB}{220,220,220}
\definecolor{codekw}{RGB}{0,92,197}
\definecolor{codestr}{RGB}{163,21,21}
\definecolor{codecomment}{RGB}{106,115,125}
\definecolor{codenum}{RGB}{150,150,150}
\definecolor{codefunc}{RGB}{111,66,193}
\lstdefinestyle{python}{
    language=Python,
    basicstyle=\ttfamily\footnotesize,
    keywordstyle=\color{codekw}\bfseries,
    keywordstyle=[2]\color{codefunc},
    morekeywords=[2]{jnp,sin,cos,array,wrap_theta},
    emph={self},
    emphstyle=\color{codekw}\itshape,
    stringstyle=\color{codestr},
    commentstyle=\color{codecomment}\itshape,
    backgroundcolor=\color{codebg},
    rulecolor=\color{codeframe},
    frame=single,
    framesep=4pt,
    xleftmargin=1.5em,
    framexleftmargin=1.5em,
    numbers=left,
    numberstyle=\tiny\color{codenum},
    numbersep=6pt,
    showstringspaces=false,
    breaklines=true,
    columns=fullflexible,
    keepspaces=true,
    tabsize=4,
    aboveskip=0.75em,
    belowskip=0.75em,
}

\usetikzlibrary{patterns,patterns.meta,arrows,arrows.meta,calc,shapes,shadows,decorations.pathmorphing,decorations.pathreplacing,automata,shapes.multipart,positioning,shapes.geometric,fit,circuits,trees,shapes.gates.logic.US,fit, shadows.blur,shapes.symbols,intersections}

\usepackage{fontawesome5}

\usepackage{dirtree}

\usepackage{tabularx} %
\usepackage{enumitem}

\usepackage{booktabs}
\usepackage{threeparttable}
\usepackage{multirow}

\newtheorem{theorem}{Theorem}
\newtheorem{assumption}{Assumption}
\newtheorem{definition}{Definition}

\newtheorem{problem}{Problem}
\newtheorem{remark}{Remark}

  {\list{}{\leftmargin=0.16in\rightmargin=0.16in}\item[]}%
  {\endlist}

%% file: include/macros.tex
\newcommand{\system}{\mathbf{S}}
\newcommand{\X}{\mathcal{X}}
\newcommand{\Xc}{\X_{\mathrm{C}}}
\newcommand{\U}{\mathcal{U}}
\newcommand{\Y}{\mathcal{Y}}
\newcommand*{\policy}{\ensuremath{\pi}}

\newcommand{\satprob}{\mathrm{Pr}}
\newcommand{\PathsFin}{\ensuremath{\mathsf{paths}^\text{fin}}}
\newcommand{\PathsInf}{\ensuremath{\mathsf{paths}^\text{inf}}}

\newcommand{\noise}{{\eta}}
\newcommand{\Noise}{\mathcal{W}}

\newcommand{\Dataset}{\mathcal{D}_N}

\newcommand*{\States}{\ensuremath{\mathcal{S}}}
\newcommand*{\Actions}{\ensuremath{\mathcal{A}}}
\newcommand*{\Sinit}{\ensuremath{s_0}}

\newcommand*{\Pl}{\ensuremath{\check{P}}}
\newcommand*{\Pu}{\ensuremath{\hat{P}}}
\newcommand*{\labelfunc}{\ensuremath{L}}
\newcommand*{\Labels}{\ensuremath{\mathcal{Y}}}

\newcommand*{\policyIMDP}{\ensuremath{\sigma}}

\newcommand*{\rmdp}{\ensuremath{\mathbf{M}}}
\newcommand*{\RMDP}{\ensuremath{\tuple{\States,\Sinit,\Actions,\cP,\Y,\labelfunc}}}
\newcommand{\cP}{\mathcal{P}}

\newcommand*{\StatesDFA}{\ensuremath{\mathcal{Q}}}

\newcommand*{\PDFA}{\ensuremath{\delta}}

\newcommand*{\horizon}{\ensuremath{K}}

\newcommand{\True}{\ensuremath{\mathsf{True}}}

\newcommand*{\Nat}{\mathbb{N}}

\newcommand*{\RR}{\mathbb{R}}
\newcommand*{\Real}{\RR} %

\newcommand*{\NN}{\mathbb{N}}

\newcommand*{\tuple}[1]{(#1)}

\newcommand*{\Prob}{\mathbb{P}}

\newcommand*{\distr}[1]{\Delta(#1)}

\newcommand{\Finally}{\lozenge}

\newcommand{\Next}{\bigcirc}
\newcommand{\UUntil}{\ensuremath{\,\mathsf U \,}}

\definecolor{cmdbg}{RGB}{238,238,238}
\ExplSyntaxOn
\NewDocumentCommand{\cmd}{v}{
    \tikz[baseline]
    \node[anchor=base, fill=cmdbg, rounded~corners=1.5pt, inner~xsep=2pt,
          inner~ysep=1pt, text~height=1.6ex, text~depth=0.45ex, font=\ttfamily]
    { \str_map_inline:nn {#1} { \str_if_eq:nnTF {##1} {~} { \nobreakspace } { ##1 } \kern 0pt } };
}
\ExplSyntaxOff

%% file: include/colors.tex
\definecolor{red}{rgb}{0.745,0.192,0.102}
\definecolor{darkgreen}{RGB}{34,161,55}
\definecolor{darkblue}{RGB}{0,33,71}
\definecolor{ruhuisstijlrood}{rgb}{0.745,0.192,0.102}
\definecolor{ruhuisstijlzwart}{rgb}{0,0,0}
\definecolor{ruhuisstijlwit}{rgb}{0.98,0.98,0.98}

\definecolor{darkred}{rgb}{0.560784314,0.125490196,0.0666666667}

\definecolor{color1}{RGB}{55,126,184} %
\definecolor{color2}{RGB}{228,26,28} %
\definecolor{color3}{RGB}{77,175,74} %
\definecolor{color4}{RGB}{152,78,163} %
\definecolor{color5}{RGB}{255,127,0} %
\definecolor{color6}{rgb}{0.5, 1.0, 0.83} %
\definecolor{color7}{rgb}{1.0, 0.0, 1.0} %
\definecolor{color8}{rgb}{0.66, 0.66, 0.66} %

\colorlet{plot1}{color1}
\colorlet{plot2}{color2}
\colorlet{plot3}{color3}
\colorlet{plot4}{color4}
\colorlet{plot5}{color5}

%% file: figures/intro_samples.tex
\begin{tikzpicture}[
    sample/.style={blue!60!black, line width=0.6pt, -{Stealth[length=3pt]}},
]
    \path[overlay] (0,0) coordinate (base);

    \node[inner sep=0pt] (drone) at ($(base) + (0.8,3.4cm)$)
        {\phantom{\includegraphics[width=1.2cm]{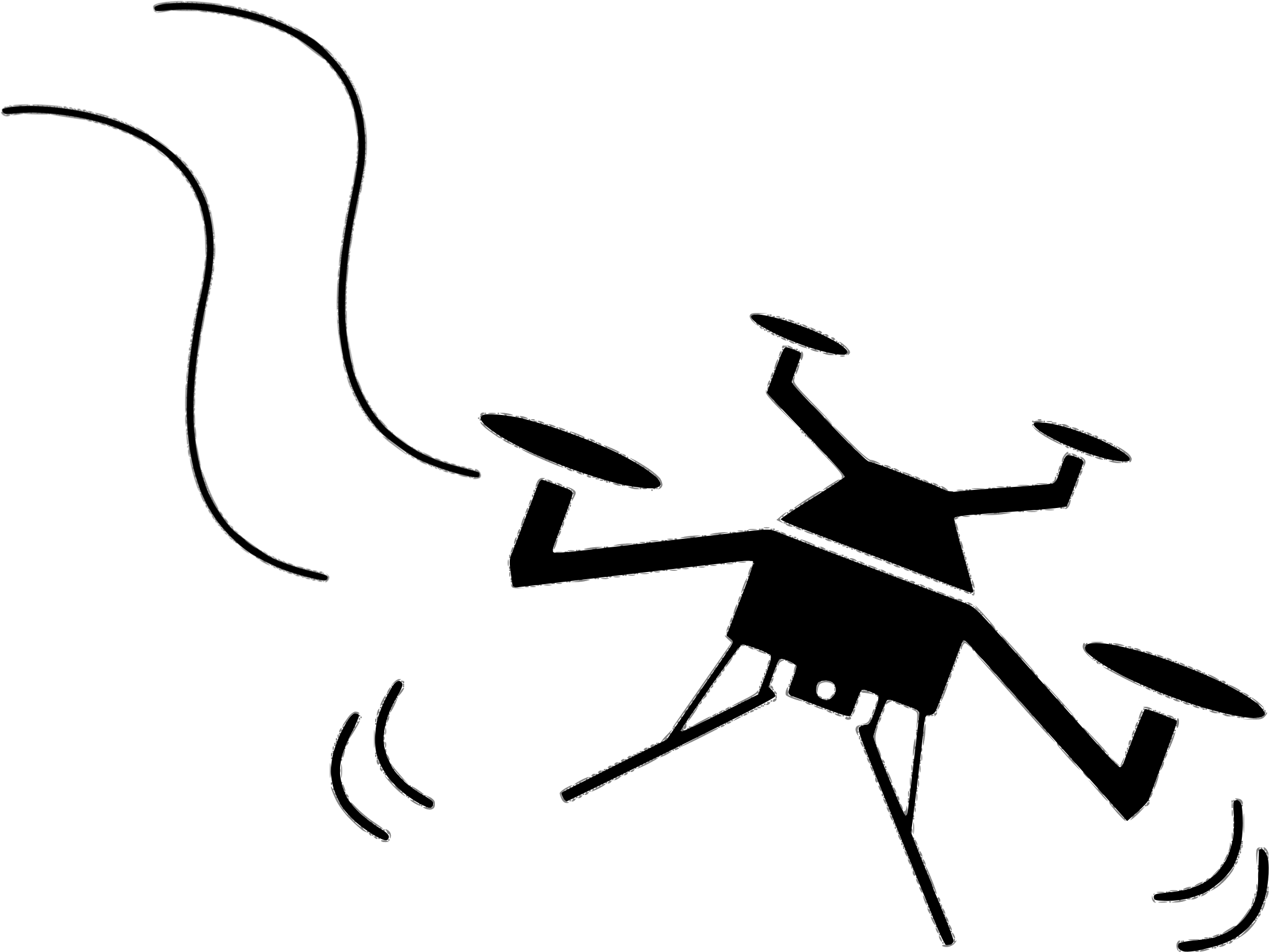}}};

    \node (box) at ($(base) + (3.3,2.5)$) {{\includegraphics[height=0.5cm]{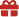}}};
    \node (goal) at ($(base) + (0.6,1.3)$) {{\includegraphics[height=0.5cm]{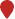}}};
    \node (tree1) at ($(base) + (2.6,3.4)$) {{\includegraphics[height=0.7cm]{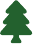}}};
    \node (tree2) at ($(base) + (3.1,3.5)$) {{\includegraphics[height=0.75cm]{figures/icons/tree}}};
    \node (tree3) at ($(base) + (1,2.05)$) {{\includegraphics[height=0.85cm]{figures/icons/tree}}};
    \node (tree4) at ($(base) + (1.4,2.4)$) {{\includegraphics[height=0.7cm]{figures/icons/tree}}};
    \node (tree5) at ($(base) + (1.9,2.1)$) {{\includegraphics[height=0.75cm]{figures/icons/tree}}};

    \foreach \xa/\ya/\xb/\yb/\bend in {
        0.35/3.55/0.85/3.75/left,
        1.25/3.25/1.75/3.45/right,
        0.35/2.95/0.80/2.75/left,
        2.00/3.05/2.20/2.60/right,
        2.50/2.85/2.85/2.50/left,
        2.45/1.90/2.95/1.80/right,
        3.35/1.95/3.50/1.45/left,
        2.15/1.35/2.65/1.15/left,
        1.15/1.45/1.55/1.10/right,
        0.20/2.20/0.45/2.55/right%
    } {
        \fill[blue!60!black] ($(base) + (\xa,\ya)$) circle (1.6pt);
        \draw[sample] ($(base) + (\xa,\ya)$) to[bend \bend=25] ($(base) + (\xb,\yb)$);
    }
\end{tikzpicture}

%% file: figures/intro_automaton.tex
\begin{tikzpicture}[
    >={Stealth[length=3pt]},
    every state/.style={draw, thick, circle, minimum size=4mm, inner sep=0pt, font=\scriptsize},
    edge/.style={->, thick},
    icon/.style={inner sep=1pt},
]
    \node[state, initial, initial text={}, initial distance=2mm] (q0) at (0,0) {$q_0$};
    \node[state] (q1) at (0.75,0) {$q_1$};
    \node[state, accepting] (q2) at (1.5,0) {$q_2$};
    \node[state] (qt) at (0.75,-0.8) {$q_\bot$};

    \draw[edge] (q0) -- node[icon, above] {\includegraphics[height=2.2mm]{figures/icons/gift}} (q1);
    \draw[edge] (q1) -- node[icon, above] {\includegraphics[height=2.2mm]{figures/icons/mapmark}} (q2);
    \draw[edge] (q0) -- node[icon, below left] {\includegraphics[height=2.2mm]{figures/icons/tree}} (qt);
    \draw[edge] (q1) -- node[icon, right] {\includegraphics[height=2.2mm]{figures/icons/tree}} (qt);
\end{tikzpicture}

%% file: figures/intro_trajectory.tex
\begin{tikzpicture}
    \path[overlay] (0,0) coordinate (base);

    \node[inner sep=0pt] (drone) at ($(base) + (0.8,3.4cm)$)
        {\includegraphics[width=1.2cm]{figures/icons/drone.png}};

    \node (box) at ($(base) + (3.3,2.5)$) {{\includegraphics[height=0.5cm]{figures/icons/gift}}};
    \node (goal) at ($(base) + (0.6,1.3)$) {{\includegraphics[height=0.5cm]{figures/icons/mapmark}}};
    \node (tree1) at ($(base) + (2.6,3.4)$) {{\includegraphics[height=0.7cm]{figures/icons/tree}}};
    \node (tree2) at ($(base) + (3.1,3.5)$) {{\includegraphics[height=0.75cm]{figures/icons/tree}}};
    \node (tree3) at ($(base) + (1,2.05)$) {{\includegraphics[height=0.85cm]{figures/icons/tree}}};
    \node (tree4) at ($(base) + (1.4,2.4)$) {{\includegraphics[height=0.7cm]{figures/icons/tree}}};
    \node (tree5) at ($(base) + (1.9,2.1)$) {{\includegraphics[height=0.75cm]{figures/icons/tree}}};

    \draw [red!80!black, ->, line width=0.6pt] plot [smooth, tension=0.9] coordinates { ($(base) + (1.4,3.3)$) ($(base) + (1.9,3.2)$) ($(base) + (2.2,2.6)$) (box.west) };
    \draw [red!80!black, ->, line width=0.6pt] plot [smooth, tension=0.9] coordinates { (box.south) ($(base) + (3,1.5)$) ($(base) + (1.8,1.2)$) (goal.east) };
\end{tikzpicture}

%% file: figures/tool_overview.tex
\def\boxradius{2pt}
\begin{tikzpicture}[
    font=\small,
    every node/.style={align=center},
    io/.style={text width=2.8cm},
    external/.style={draw, thick, rounded corners=\boxradius, minimum width=2.6cm, minimum height=6mm},
    stepbox/.style={draw=darkgreen!75!black, fill=darkgreen!10, thick, rounded corners=\boxradius, minimum height=6mm},
    halfstep/.style={stepbox, text width=2.1cm, minimum width=2.4cm, minimum height=9mm},
    fullstep/.style={stepbox, text width=4.9cm, minimum width=5.2cm},
    toolbox/.style={draw=darkgreen!75!black, thick, rounded corners=\boxradius, inner sep=3mm},
    arrow/.style={-{Stealth}, thick},
]
    \node[font=\large] (title) {\textbf\tool};
    \node[halfstep, below=1mm of title, xshift=-1.4cm] (abs) {Abstraction\\(IMDP or 2IMDP)};
    \node[halfstep, right=4mm of abs] (dfa) {DFA construction};
    \coordinate (row) at ($(abs.south)!0.5!(dfa.south)$);

    \node[fullstep, below=3mm of row] (dp) {Robust dynamic programming};
    \draw[arrow] (abs) -- (abs |- dp.north);
    \draw[arrow] (dfa) -- (dfa |- dp.north);

    \node[stepbox, text width=2.1cm, minimum width=2.4cm, below=3mm of dp.south -| abs] (ref) {Policy refinement};
    \draw[arrow] (abs |- dp.south) -- (ref);
    \node[toolbox, fit={($(title.north)+(0,-2mm)$) (abs) (dfa) (dp) (ref)}] (tool) {};

    \node[io, above=3mm of tool.north, xshift=-1.4cm] (data) {Dataset $\mathcal{D}_N$};
    \node[io, above=3mm of tool.north, xshift=1.4cm] (spec) {scLTL specification $\varphi$};
    \node[external, above=4mm of data] (sim) {System simulator};
    \draw[arrow] (sim) -- (data);
    \draw[arrow] (data) -- (abs);
    \draw[arrow] (spec) -- (dfa);
    \node[io, text width=1.6cm, left=4mm of tool.west |- abs] (beta) {Confidence\\parameter $\beta$};
    \draw[arrow] (beta) -- (abs);

    \node[io, below=4mm of tool.south, xshift=-1.4cm] (policy) {Finite-memory policy $\policy$};
    \node[io, below=4mm of tool.south, xshift=1.4cm] (bound) {Certified bound on satisfaction probability};
    \draw[arrow] (ref) -- (policy);
    \draw[arrow] (dfa |- dp.south) -- (bound);
\end{tikzpicture}

%% file: figures/two_phase_sampling.tex
\begin{tikzpicture}[
    scale=0.45,
    font=\small,
    grid/.style={gray!50, thin},
    source/.style={fill=gray!35},
    candidate/.style={fill=color1!15},
    sample/.style={draw=black!70, fill=black!15, line width=0.4pt},
    inside/.style={draw=color1, fill=color1!35, line width=0.4pt},
    leaving/.style={draw=color2, fill=color2!25, line width=0.4pt},
    bbox/.style={thick, dashed},
    drift/.style={-{Stealth[length=3pt]}, semithick, gray!70!black},
]
    \def\hw{0.18}

    \def\source{
        \fill[source] (0, 0) rectangle (1, 1);
        \node at (0.5, 0.5) {$R_s$};
        \draw[drift] (1.05, 1.05) to[bend left=15] node[pos=0.45, left, fill=white, inner sep=1pt, xshift=-0.7mm] {$u$} (2.4, 2.4);
    }

    \begin{scope}
        \fill[candidate] (2, 2) rectangle (5, 5);
        \source
        \draw[grid] (0, 0) grid (6, 6);
        \foreach \x/\y in {2.7/3.5, 3.2/2.7, 3.75/3.4, 3.4/4.0, 2.9/3.0, 4.05/2.8} {
            \draw[sample] (\x-\hw, \y-\hw) rectangle (\x+\hw, \y+\hw);
        }
        \draw[bbox] (2.52, 2.52) rectangle (4.23, 4.18);
        \draw[thick] (0, 0) rectangle (6, 6);
        \node[below] at (3, 0) {(a) First batch};
    \end{scope}

    \begin{scope}[xshift=7cm]
        \fill[candidate] (2, 2) rectangle (5, 5);
        \source
        \draw[grid] (0, 0) grid (6, 6);
        \draw[thick, color1] (2, 2) rectangle (5, 5);
        \foreach \x/\y in {2.6/2.9, 3.1/3.6, 3.5/2.6, 3.9/3.9, 3.3/3.15, 4.3/3.3, 2.8/4.2, 3.7/4.5} {
            \draw[inside] (\x-\hw, \y-\hw) rectangle (\x+\hw, \y+\hw);
        }
        \foreach \x/\y in {1.95/3.5, 4.9/2.7, 3.4/5.05} {
            \draw[leaving] (\x-\hw, \y-\hw) rectangle (\x+\hw, \y+\hw);
        }
        \draw[thick] (0, 0) rectangle (6, 6);
        \node[below] at (3, 0) {(b) Second batch};
    \end{scope}
\end{tikzpicture}

%% file: figures/codebase.tex
\newcommand{\dtfolder}[1]{\textcolor{black!70}{\faFolder}\,\textbf{#1}}
\newcommand{\dtfile}[1]{\textcolor{black!45}{\faFile*[regular]}\,#1}
\newcommand{\dtsec}[1]{\textcolor{black!55}{(#1)}}
\renewcommand*\DTstylecomment{\rmfamily\footnotesize}
\renewcommand*\DTstyle{\ttfamily\footnotesize}
\setlength{\DTbaselineskip}{8pt}
\DTsetlength{0.2em}{0.9em}{0.2em}{0.4pt}{1.6pt}
\dirtree{%
.1 \dtfolder{darsy/}.
.2 \dtfile{RunIMDP.py}\DTcomment{entry point for IMDP abstraction}.
.2 \dtfile{Run2IMDP.py}\DTcomment{entry point for 2IMDP abstraction}.
.2 \dtfolder{benchmarks/}\DTcomment{benchmark definitions \dtsec{\cref{subsec:models}}}.
.3 \dtfile{Pendulum.py}\textrm{\,(etc.)}\DTcomment{one file per benchmark}.
.3 \dtfile{models.py}\DTcomment{simulators (of dynamics $f$) for all benchmarks}.
.2 \dtfolder{core/}.
.3 \dtfile{pipeline.py}\DTcomment{high-level pipeline for both entry points}.
.3 \dtfolder{abstraction/}.
.4 \dtfile{partition.py}\DTcomment{label-preserving grid \dtsec{\cref{subsec:specifications}}}.
.4 \dtfolder{imdp/}.
.5 \dtfile{forward\_reachability.py}\DTcomment{successors \dtsec{Rem.~\ref{rem:two_step}}}.
.5 \dtfile{probability\_intervals.py}\DTcomment{intervals \dtsec{\cref{sec:abstraction}}}.
.5 \dtfile{robust\_DP.py}\DTcomment{robust DP \dtsec{\cref{sec:synthesis}}}.
.3 \dtfolder{automaton/}\DTcomment{scLTL $\to$ DFA via Spot \dtsec{\cref{sec:synthesis}}}.
.3 \dtfolder{validate/}\DTcomment{Monte Carlo runs of Alg.~\ref{alg:closed_loop}}.
.3 \dtfolder{plotting/}\DTcomment{heatmaps, trajectories, GIFs, etc.}.
}

%% file: figures/partition.tex
\begin{tikzpicture}[
    scale=0.6,
    font=\small,
    nominal/.style={gray!50, thin},
    removed/.style={gray!70, semithick, dotted},
    added/.style={color1, semithick, dash pattern=on 2.5pt off 2pt},
    region/.style={thick},
    regionlabel/.style={inner sep=1pt},
]
    \coordinate (goal-lb) at (3, 3.5);   \coordinate (goal-ub) at (5, 5);
    \coordinate (unsafe-lb) at (1, 1);   \coordinate (unsafe-ub) at (2.5, 3);
    \def\xkept{1, 3, 4, 5}       \def\xremoved{2}  \def\xadded{1.75, 2.5}
    \def\ykept{1, 2, 3, 5}       \def\yremoved{4}  \def\yadded{3.5, 4.25}

    \fill[darkgreen!20] (goal-lb) rectangle (goal-ub);
    \fill[red!20] (unsafe-lb) rectangle (unsafe-ub);

    \foreach \x in \xkept { \draw[nominal] (\x, 0) -- (\x, 6); }
    \foreach \y in \ykept { \draw[nominal] (0, \y) -- (6, \y); }
    \foreach \x in \xremoved { \draw[removed] (\x, 0) -- (\x, 6); }
    \foreach \y in \yremoved { \draw[removed] (0, \y) -- (6, \y); }

    \foreach \x in \xadded { \draw[added] (\x, 0) -- (\x, 6); }
    \foreach \y in \yadded { \draw[added] (0, \y) -- (6, \y); }

    \draw[region, darkgreen] (goal-lb) rectangle (goal-ub);
    \draw[region, red] (unsafe-lb) rectangle (unsafe-ub);
    \node[regionlabel, darkgreen!70!black, fill=darkgreen!20] at (4, 4.625) {goal};
    \node[regionlabel, red!70!black, fill=red!20] at (1.75, 2.5) {unsafe};

    \draw[thick] (0, 0) rectangle (6, 6);
    \node[below] at (3, 0) {$x_1$};
    \node[left] at (0, 3) {$x_2$};

    \begin{scope}[shift={(6.5, 3)}, font=\footnotesize]
        \draw[nominal] (0, 0.8) -- (0.6, 0.8) node[right, black] {nominal grid};
        \draw[added] (0, 0) -- (0.6, 0) node[right, black] {added slice};
        \draw[removed] (0, -0.8) -- (0.6, -0.8) node[right, black] {removed slice};
    \end{scope}
\end{tikzpicture}

%% file: tables/table_main_extended.tex
\begin{tabular}{llrrr | rrrr | rrrr | rrrr}
\toprule
& & & & \multicolumn{1}{r}{} & \multicolumn{4}{c}{1\,000 samples} & \multicolumn{4}{c}{10\,000 samples} & \multicolumn{4}{c}{100\,000 samples} \\
\cmidrule(lr){6-9} \cmidrule(lr){10-13} \cmidrule(lr){14-17}
\textbf{Benchmark} & \textbf{Method} & $|S|$ & $|A|$ & \#Trans. & Abs.\ [s] & Syn.\ [s] & $\lambda$ & Emp. & Abs.\ [s] & Syn.\ [s] & $\lambda$ & Emp. & Abs.\ [s] & Syn.\ [s] & $\lambda$ & Emp. \\
\midrule
\multirow{2}{*}{\begin{tabular}{@{}l@{}}\textbf{Temp1D}\\\textit{(reach-avoid)}\end{tabular}}
 & IMDP & $1\,002$ & $2$ & $7.41\cdot 10^{5}$ & $\boldsymbol{1.5}$ & $\boldsymbol{0.4}$ & $0.000$ & $0.000$ & $\boldsymbol{6.9}$ & $\boldsymbol{0.6}$ & $0.000$ & $0.000$ & $\boldsymbol{84.3}$ & $\boldsymbol{1.4}$ & $0.000$ & $0.994$ \\
 & 2IMDP & $1\,002$ & $2$ & $9.87\cdot 10^{5}$ & $1.9$ & $1.0$ & $0.000$ & $0.996$ & $9.4$ & $4.2$ & $\boldsymbol{0.251}$ & $0.996$ & $107.9$ & $2.7$ & $\boldsymbol{0.738}$ & $0.996$ \\
\midrule
\multirow{2}{*}{\begin{tabular}{@{}l@{}}\textbf{Temp1D}\\\textit{(safety)}\end{tabular}}
 & IMDP & $102$ & $2$ & $7.88\cdot 10^{3}$ & $\boldsymbol{0.8}$ & $\boldsymbol{0.3}$ & $0.021$ & $0.993$ & $\boldsymbol{1.0}$ & $\boldsymbol{0.3}$ & $0.302$ & $0.993$ & $\boldsymbol{2.2}$ & $\boldsymbol{0.3}$ & $0.548$ & $0.993$ \\
 & 2IMDP & $102$ & $2$ & $1.05\cdot 10^{4}$ & $0.9$ & $\boldsymbol{0.3}$ & $\boldsymbol{0.246}$ & $0.993$ & $1.1$ & $\boldsymbol{0.3}$ & $\boldsymbol{0.696}$ & $0.993$ & $2.6$ & $\boldsymbol{0.3}$ & $\boldsymbol{0.800}$ & $0.993$ \\
\midrule
\multirow{2}{*}{\begin{tabular}{@{}l@{}}\textbf{Temp2D}\\\textit{(safety)}\end{tabular}}
 & IMDP & $145$ & $4$ & $2.13\cdot 10^{4}$ & $\boldsymbol{1.2}$ & $\boldsymbol{0.3}$ & $\boldsymbol{0.768}$ & $1.000$ & $\boldsymbol{1.3}$ & $\boldsymbol{0.3}$ & $\boldsymbol{0.973}$ & $1.000$ & $\boldsymbol{2.4}$ & $\boldsymbol{0.3}$ & $\boldsymbol{0.997}$ & $1.000$ \\
 & 2IMDP & $145$ & $4$ & $2.36\cdot 10^{4}$ & $1.3$ & $\boldsymbol{0.3}$ & $0.767$ & $1.000$ & $1.5$ & $\boldsymbol{0.3}$ & $\boldsymbol{0.973}$ & $1.000$ & $3.1$ & $\boldsymbol{0.3}$ & $\boldsymbol{0.997}$ & $1.000$ \\
\midrule
\multirow{2}{*}{\begin{tabular}{@{}l@{}}\textbf{Temp3D}\\\textit{(safety)}\end{tabular}}
 & IMDP & $1\,729$ & $8$ & $3.00\cdot 10^{6}$ & $\boldsymbol{2.2}$ & $\boldsymbol{0.6}$ & $\boldsymbol{0.712}$ & $1.000$ & $\boldsymbol{4.7}$ & $\boldsymbol{0.6}$ & $\boldsymbol{0.966}$ & $1.000$ & $\boldsymbol{43.8}$ & $\boldsymbol{0.6}$ & $\boldsymbol{0.996}$ & $1.000$ \\
 & 2IMDP & $1\,729$ & $8$ & $3.11\cdot 10^{6}$ & $2.5$ & $2.1$ & $\boldsymbol{0.712}$ & $1.000$ & $7.2$ & $2.0$ & $\boldsymbol{0.966}$ & $1.000$ & $71.8$ & $2.2$ & $\boldsymbol{0.996}$ & $1.000$ \\
\midrule
\multirow{2}{*}{\begin{tabular}{@{}l@{}}\textbf{Temp4D}\\\textit{(safety)}\end{tabular}}
 & IMDP & $20\,737$ & $16$ & $4.30\cdot 10^{8}$ & $\boldsymbol{18.7}$ & $\boldsymbol{69.6}$ & $0.617$ & $1.000$ & $\boldsymbol{103.8}$ & $\boldsymbol{71.4}$ & $0.958$ & $1.000$ & $\boldsymbol{1381.0}$ & $\boldsymbol{70.8}$ & $\boldsymbol{0.995}$ & $1.000$ \\
 & 2IMDP & $20\,737$ & $16$ & $4.36\cdot 10^{8}$ & $25.5$ & $490.6$ & $\boldsymbol{0.633}$ & $1.000$ & $171.3$ & $504.0$ & $\boldsymbol{0.959}$ & $1.000$ & $4988.3$ & $495.5$ & $\boldsymbol{0.995}$ & $1.000$ \\
\midrule
\multirow{2}{*}{\begin{tabular}{@{}l@{}}\textbf{DoubleInt}\\\textit{(reach-avoid)}\end{tabular}}
 & IMDP & $2\,705$ & $21$ & $1.19\cdot 10^{6}$ & $\boldsymbol{1.9}$ & $\boldsymbol{0.4}$ & $\boldsymbol{0.917}$ & $1.000$ & $\boldsymbol{6.4}$ & $\boldsymbol{0.4}$ & $\boldsymbol{0.991}$ & $1.000$ & $\boldsymbol{41.8}$ & $\boldsymbol{0.4}$ & $\boldsymbol{0.999}$ & $1.000$ \\
 & 2IMDP & $2\,705$ & $21$ & $1.42\cdot 10^{6}$ & $2.5$ & $0.8$ & $0.916$ & $1.000$ & $12.0$ & $0.8$ & $\boldsymbol{0.991}$ & $1.000$ & $113.4$ & $0.9$ & $\boldsymbol{0.999}$ & $1.000$ \\
\midrule
\multirow{2}{*}{\begin{tabular}{@{}l@{}}\textbf{DoubleInt}\\\textit{(scLTL $\varphi_2$)}\end{tabular}}
 & IMDP & $2\,809$ & $21$ & $1.24\cdot 10^{6}$ & $\boldsymbol{1.8}$ & $\boldsymbol{0.9}$ & $\boldsymbol{0.794}$ & $1.000$ & $\boldsymbol{6.5}$ & $\boldsymbol{1.0}$ & $\boldsymbol{0.978}$ & $1.000$ & $\boldsymbol{42.8}$ & $\boldsymbol{0.8}$ & $\boldsymbol{0.997}$ & $1.000$ \\
 & 2IMDP & $2\,809$ & $21$ & $1.36\cdot 10^{6}$ & $2.4$ & $2.6$ & $\boldsymbol{0.794}$ & $1.000$ & $11.9$ & $2.6$ & $\boldsymbol{0.978}$ & $1.000$ & $115.0$ & $2.5$ & $\boldsymbol{0.997}$ & $1.000$ \\
\midrule
\multirow{2}{*}{\begin{tabular}{@{}l@{}}\textbf{MountainCar}\\\textit{(reach-avoid)}\end{tabular}}
 & IMDP & $250\,501$ & $3$ & $3.23\cdot 10^{7}$ & $\boldsymbol{27.8}$ & $\boldsymbol{242.2}$ & $0.055$ & $1.000$ & $\boldsymbol{93.7}$ & $\boldsymbol{210.7}$ & $0.770$ & $1.000$ & $\boldsymbol{1037.1}$ & $\boldsymbol{202.3}$ & $\boldsymbol{0.974}$ & $1.000$ \\
 & 2IMDP & $250\,501$ & $3$ & $3.53\cdot 10^{7}$ & $28.9$ & $1112.3$ & $\boldsymbol{0.059}$ & $1.000$ & $175.1$ & $1118.3$ & $\boldsymbol{0.772}$ & $1.000$ & $1750.0$ & $1159.1$ & $\boldsymbol{0.974}$ & $1.000$ \\
\midrule
\multirow{2}{*}{\begin{tabular}{@{}l@{}}\textbf{Pendulum-T}\\\textit{(reach-avoid)}\end{tabular}}
 & IMDP & $40\,201$ & $3$ & $5.19\cdot 10^{6}$ & $\boldsymbol{5.6}$ & $\boldsymbol{10.1}$ & $0.416$ & $1.000$ & $\boldsymbol{18.5}$ & $\boldsymbol{10.3}$ & $0.925$ & $1.000$ & $\boldsymbol{212.9}$ & $\boldsymbol{8.8}$ & $\boldsymbol{0.992}$ & $1.000$ \\
 & 2IMDP & $40\,201$ & $3$ & $5.67\cdot 10^{6}$ & $5.8$ & $57.6$ & $\boldsymbol{0.419}$ & $1.000$ & $32.2$ & $48.4$ & $\boldsymbol{0.926}$ & $1.000$ & $324.1$ & $43.5$ & $\boldsymbol{0.992}$ & $1.000$ \\
\midrule
\multirow{2}{*}{\begin{tabular}{@{}l@{}}\textbf{Pendulum-G}\\\textit{(reach-avoid)}\end{tabular}}
 & IMDP & $40\,201$ & $3$ & $2.36\cdot 10^{7}$ & $\boldsymbol{7.6}$ & $\boldsymbol{3.3}$ & $0.000$ & $0.108$ & $\boldsymbol{22.2}$ & $\boldsymbol{69.2}$ & $0.000$ & $0.967$ & $\boldsymbol{224.5}$ & $\boldsymbol{166.3}$ & $0.622$ & $1.000$ \\
 & 2IMDP & $40\,201$ & $3$ & $2.60\cdot 10^{7}$ & $8.8$ & $23.7$ & $0.000$ & $0.997$ & $38.4$ & $338.5$ & $\boldsymbol{0.146}$ & $1.000$ & $349.6$ & $250.8$ & $\boldsymbol{0.743}$ & $1.000$ \\
\midrule
\multirow{2}{*}{\begin{tabular}{@{}l@{}}\textbf{Dubins2D}\\\textit{(scLTL $\varphi_1$)}\end{tabular}}
 & IMDP & $901$ & $40$ & $9.36\cdot 10^{5}$ & $\boldsymbol{1.3}$ & $\boldsymbol{1.0}$ & $0.474$ & $1.000$ & $\boldsymbol{3.1}$ & $\boldsymbol{0.9}$ & $0.937$ & $1.000$ & $\boldsymbol{25.0}$ & $\boldsymbol{0.9}$ & $\boldsymbol{0.993}$ & $1.000$ \\
 & 2IMDP & $901$ & $40$ & $9.72\cdot 10^{5}$ & $\boldsymbol{1.3}$ & $3.0$ & $\boldsymbol{0.495}$ & $1.000$ & $3.4$ & $2.7$ & $\boldsymbol{0.939}$ & $1.000$ & $28.1$ & $2.9$ & $\boldsymbol{0.993}$ & $1.000$ \\
\midrule
\multirow{2}{*}{\begin{tabular}{@{}l@{}}\textbf{Dubins3D}\\\textit{(reach-avoid)}\end{tabular}}
 & IMDP & $127\,501$ & $9$ & $1.16\cdot 10^{8}$ & $\boldsymbol{48.4}$ & $\boldsymbol{39.3}$ & $0.356$ & $1.000$ & $\boldsymbol{294.6}$ & $\boldsymbol{36.2}$ & $\boldsymbol{0.918}$ & $1.000$ & $\boldsymbol{3272.0}$ & $\boldsymbol{34.1}$ & $\boldsymbol{0.991}$ & $1.000$ \\
 & 2IMDP & $127\,501$ & $9$ & $1.18\cdot 10^{8}$ & $66.4$ & $559.6$ & $\boldsymbol{0.367}$ & $1.000$ & $501.4$ & $502.1$ & $\boldsymbol{0.918}$ & $1.000$ & $5534.3$ & $465.6$ & $\boldsymbol{0.991}$ & $1.000$ \\
\midrule
\multirow{2}{*}{\begin{tabular}{@{}l@{}}\textbf{Dubins3D}\\\textit{(scLTL $\varphi_1$)}\end{tabular}}
 & IMDP & $127\,501$ & $9$ & $1.16\cdot 10^{8}$ & $\boldsymbol{48.4}$ & $\boldsymbol{175.4}$ & $0.330$ & $1.000$ & $\boldsymbol{291.8}$ & $\boldsymbol{155.1}$ & $\boldsymbol{0.915}$ & $1.000$ & $\boldsymbol{3364.2}$ & $\boldsymbol{146.9}$ & $\boldsymbol{0.991}$ & $1.000$ \\
 & 2IMDP & $127\,501$ & $9$ & $1.18\cdot 10^{8}$ & $66.5$ & $906.2$ & $\boldsymbol{0.341}$ & $1.000$ & $493.9$ & $803.6$ & $\boldsymbol{0.915}$ & $1.000$ & $5547.6$ & $746.1$ & $\boldsymbol{0.991}$ & $1.000$ \\
\bottomrule
\end{tabular}%